\documentclass[%
reprint,
superscriptaddress,
amsmath,amssymb,
prx
]{revtex4-2}

\usepackage{xcolor}
\usepackage{multirow}
\usepackage{graphicx}
\usepackage{dcolumn}
\usepackage{bm}
\usepackage{hyperref}

\begin{document}

\preprint{APS/123-QED}

\title{Unconventional Coherence Peak in Cuprate Superconductors}

\author{Zheng Li}
\email{lizheng@iphy.ac.cn}
\affiliation{Beijing National Laboratory for Condensed Matter Physics and Institute of Physics, Chinese Academy of Sciences, Beijing 100190, China}
\affiliation{School of Physical Sciences, University of Chinese Academy of Sciences, Beijing 100190, China}

\author{Chao Mu}
\affiliation{Beijing National Laboratory for Condensed Matter Physics and Institute of Physics, Chinese Academy of Sciences, Beijing 100190, China}
\affiliation{School of Physical Sciences, University of Chinese Academy of Sciences, Beijing 100190, China}

\author{Pengfei Li}
\affiliation{Beijing National Laboratory for Condensed Matter Physics and Institute of Physics, Chinese Academy of Sciences, Beijing 100190, China}
\affiliation{School of Physical Sciences, University of Chinese Academy of Sciences, Beijing 100190, China}

\author{Wei Wu}
\affiliation{Beijing National Laboratory for Condensed Matter Physics and Institute of Physics, Chinese Academy of Sciences, Beijing 100190, China}

\author{Jiangping Hu}
\affiliation{Beijing National Laboratory for Condensed Matter Physics and Institute of Physics, Chinese Academy of Sciences, Beijing 100190, China}
\affiliation{New Cornerstone Science Laboratory, Beijing 100190, China.}

\author{Tao Xiang}
\affiliation{Beijing National Laboratory for Condensed Matter Physics and Institute of Physics, Chinese Academy of Sciences, Beijing 100190, China}
\affiliation{School of Physical Sciences, University of Chinese Academy of Sciences, Beijing 100190, China}

\author{Kun Jiang}
\email{jiangkun@iphy.ac.cn}
\affiliation{Beijing National Laboratory for Condensed Matter Physics and Institute of Physics, Chinese Academy of Sciences, Beijing 100190, China}
\affiliation{School of Physical Sciences, University of Chinese Academy of Sciences, Beijing 100190, China}

\author{Jianlin Luo}
\email{jlluo@iphy.ac.cn}
\affiliation{Beijing National Laboratory for Condensed Matter Physics and Institute of Physics, Chinese Academy of Sciences, Beijing 100190, China}
\affiliation{School of Physical Sciences, University of Chinese Academy of Sciences, Beijing 100190, China}


\begin{abstract}
The Hebel-Slichter coherence peak, observed in the spin-lattice relaxation rate $1/T_1$ just below the critical temperature $T_{\rm c}$, serves as a crucial experimental validation of the Bardeen-Cooper-Schrieffer pairing symmetry in conventional superconductors. However, no coherence peak in $1/T_1$ has been observed in unconventional superconductors like cuprates. In this study, an unconventional coherence peak is identified for the first time using nuclear quadrupole resonance on YBa$_2$Cu$_4$O$_8$, pointing to a distinctive pairing symmetry. The spin-lattice relaxation rate in nuclear quadrupole resonance and nuclear magnetic resonance with nuclear spin $I>1/2$ comprises the magnetic relaxation rate $1/T_{1}^{\rm mag}$, which probes magnetic fluctuations, and the quadrupole relaxation rate $1/T_{1}^{\rm quad}$, which probes charge fluctuations. By utilizing $^{63}$Cu and $^{65}$Cu isotopes, we successfully distinguish $1/T_{1}^{\rm mag}$ and $1/T_{1 }^{\rm quad}$ of YBa$_2$Cu$_4$O$_8$ and reveal the presence of the coherence peak in $1/T_{1 }^{\rm quad}$ but not in $1/T_{1}^{\rm mag}$, in contrast to conventional superconductors. Our finding demonstrates that unconventional superconductors do not exhibit a coherence peak in $1/T_{1}$ when the relaxation is due to fluctuations of the hyperfine field. Conversely, a coherence peak is expected when the relaxation is caused by electric field gradient fluctuations, due to the different coherence factors between charge and magnetic fluctuations. Our successful measurements of $1/T_{1}$ for the chains of YBa$_2$Cu$_4$O$_8$ suggest that, should the conditions for predominant quadrupole relaxation be satisfied, this phenomenon could provide a novel approach to exploring the unconventional nature of the pairing mechanism in other superconductors.

\end{abstract}


\maketitle


\section{Introduction}
Unconventional superconductors, such as $d$-wave and $s^{\pm}$-wave superconductors, are characterized by a varying sign of the superconducting gap function in momentum space, with their Cooper pairs widely believed to originate from electron-electron correlation rather than electron-phonon coupling\cite{Taillefer2012}. Therefore, unconventional superconductors exhibit distinct physical properties, such as the absence of the Hebel-Slichter coherence peak \cite{Imai1988Cu2,Imai1988Cu63,Zimmermann1989,Warren1990, Machi1991,ASAYAMA1996221,Imai1995,rigamonti_1998,Jurkutat2019}. In conventional superconductors with uniform superconducting gaps, the density of states (DOS) diverges at the gap energy, leading to a huge enhancement of spin-lattice relaxation rate $1/T_1$ just below the critical temperature $T_{\rm c}$ in nuclear magnetic resonance (NMR) experiment, called the Hebel-Slichter coherence peak\cite{Hebel1959}. Conversely, this Hebel-Slichter coherence peak is absent in unconventional superconductors\cite{Jurkutat2019}, such as \emph{d}-wave cuprate superconductors, where the DOS divergence persists. Consequently, detecting coherence peaks poses a significant challenge for unconventional superconductors. In this work, we reveal a novel unconventional coherence peak in high-temperature cuprate superconductors using nuclear quadrupole resonance (NQR), introducing a distinctive feature of unconventional superconductivity.

\begin{figure*}
\includegraphics[width=0.75\textwidth,clip]{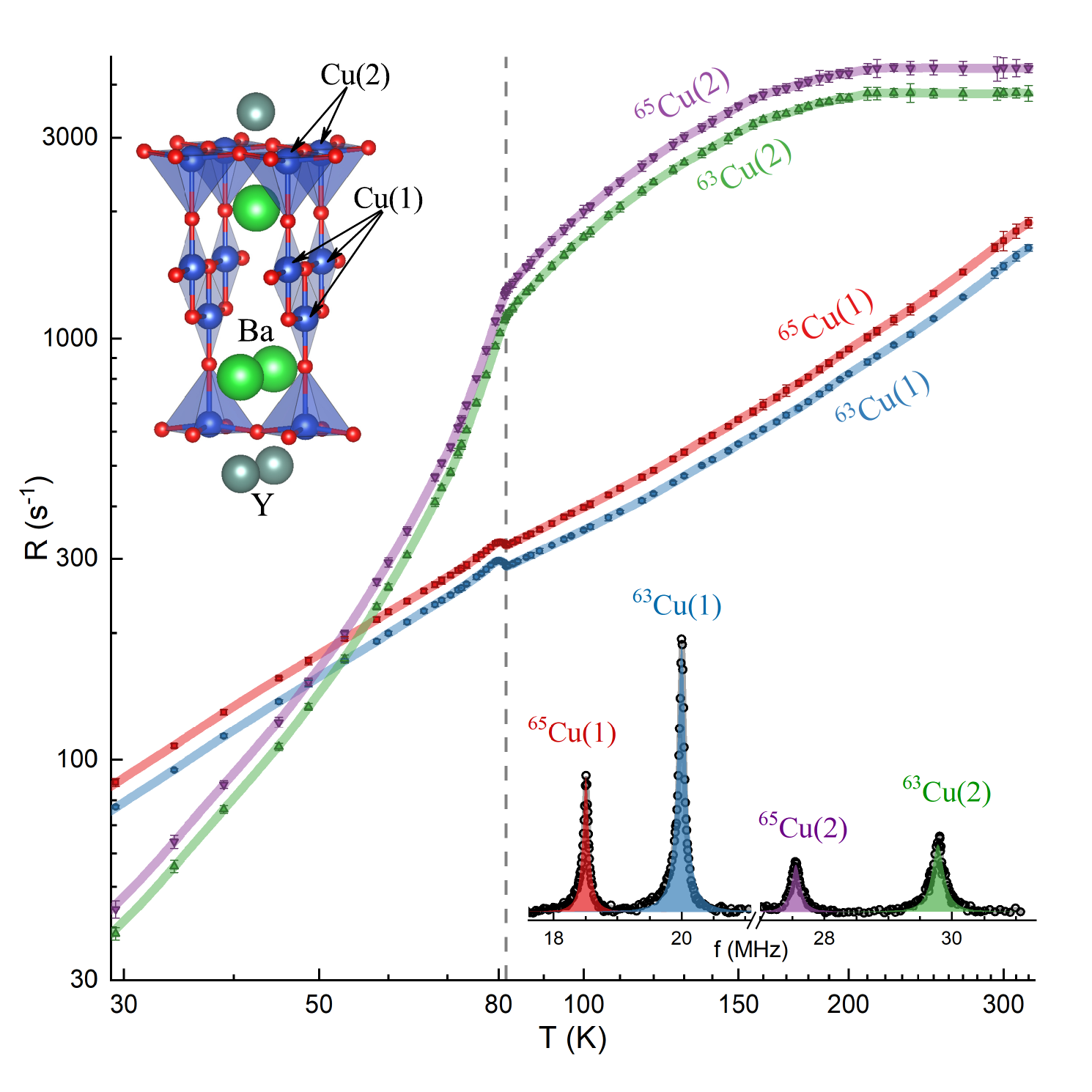}
\caption{Temperature dependence of relaxation rate $R$ of YBa$_2$Cu$_4$O$_8$. The inset crystal structure indicates the Cu(1) site and Cu(2) site. The inset spectrum shows four peaks of \textsuperscript{63}Cu and \textsuperscript{65}Cu at the Cu(1) and Cu(2) sites respectively, where $R$ is measured. The dashed line marks the superconducting transition temperature $T_{\rm c}$. $R$ of \textsuperscript{63}Cu(2) (green) and \textsuperscript{65}Cu(2) (purple) drops below $T_{\rm c}$, while $R$ of \textsuperscript{63}Cu(1) (cyan) and \textsuperscript{65}Cu(1) (red) increases a little just below $T_{\rm c}$ and forms a small peak.}\label{fig:T1}
\end{figure*}

The nucleus with the spin $I >1/2$ is nonspherical and possesses an electric quadrupole moment $Q _{\alpha \beta}$, where $\alpha$ and $\beta$ are spatial directions \emph{x}, \emph{y}, \emph{z}. This results in the nucleus having an electrostatic energy that varies depending on its orientation within an electric field gradient (EFG) $V_{\alpha \beta }  = \tfrac{{\partial ^2 V}}{{\partial \alpha \partial \beta }}$ generated by the local potential $V$ from its surrounding environments\cite{Slichter1990}. Hence, the electrostatic energy splitting between different nucleus spin $|I|$ is utilized in NQR, analogous to how NMR employs the magnetic energy splitting of each spin $I$ under a magnetic field. Furthermore, the electric quadrupole moment is a valuable tool for investigating electric field dynamics, akin to how the magnetic moment is used to study spin dynamics. These tools complement each other in exploring electromagnetic fluctuations in condensed matter\cite{Slichter2017}. A key dynamic signal is the spin-lattice relaxation rate, which describes how the nuclei arrive at their thermal equilibrium via the process of spin-lattice relaxation and is proportional to the summation of the imaginary part of the dynamical susceptibility. In the conventional $s$-wave superconducting state, the rate can be expressed as\cite{MacLaughlinDE}
\begin{equation}\label{T1s}
\frac{{T_{1 {\rm N}} }}
{{T_{1 {\rm S}} }} =  - \frac{2}
{{N_{\text{0}} ^{\text{2}} }}\int_\Delta ^\infty  {\left( {1 \pm \frac{{\Delta ^2 }}
{{E^2 }}} \right)N_{\rm S} ^2 \left( E \right)\frac{{\partial f\left( E \right)}}
{{\partial E}}\operatorname{d} E}
\end{equation}
where $T_{1 {\rm N}}$ and $T_{1 {\rm S}}$  are the relaxation times in the normal state and the superconducting state, respectively. $N_0$ is the DOS in the normal state and $N_{\rm S} (E) = N_0 E/\sqrt {E^2  - \Delta ^2 }$ for  $E>\Delta$ is the DOS in the superconducting state. $f(E)$ is the Fermi distribution function and $\Delta$ is the superconducting gap. The sign of the coherence factor $\left( {1 \pm {{\Delta ^2 } \mathord{\left/ {\vphantom {{\Delta ^2 } {E^2 }}} \right. \kern-\nulldelimiterspace} {E^2 }}} \right)$ is contingent on the nature of the perturbation causing the transition. Magnetic relaxations, being not time-reversal invariant, have $\left( {1 + {{\Delta ^2 } \mathord{\left/ {\vphantom {{\Delta ^2 } {E^2 }}} \right. \kern-\nulldelimiterspace} {E^2 }}} \right)$ which enhance the divergency of $N_{\rm S} (E)$ and the magnetic relaxation rate $1/T_{1}^{\rm mag}$ exhibits a Hebel-Slichter coherence peak in conventional superconductors\cite{MacLaughlinDE}. In contrast, quadrupole relaxations, being time-reversal invariant, have $\left( {1 - {{\Delta ^2 } \mathord{\left/ {\vphantom {{\Delta ^2 } {E^2 }}} \right. \kern-\nulldelimiterspace} {E^2 }}} \right)$ which compensate for the divergency of $N_{\rm S} (E)$ and the quadrupole relaxation rate $1/T_{1 }^{\rm quad}$ drops rapidly below $T_{\rm c}$ in conventional superconductors\cite{Wada1973,Li2016TaPdTe}. In unconventional superconductors, $1/T_{1}^{\rm mag}$ drops or decreases gradually below $T_{\rm c}$, without a coherence peak being observed\cite{Imai1988Cu2}. These phenomena contradict the logarithmic divergence of $N_{\rm S} (E)$ in cuprate superconductors\cite{Xiang2022}. Additionally, the cuprate superconductor gap sign varies in momentum space, significantly impacting the coherence factor and potentially leading to a shift of the coherence peak to $1/T_{1 }^{\rm quad}$.

\section{Result}

\begin{figure*}
\includegraphics[width=0.75\textwidth,clip]{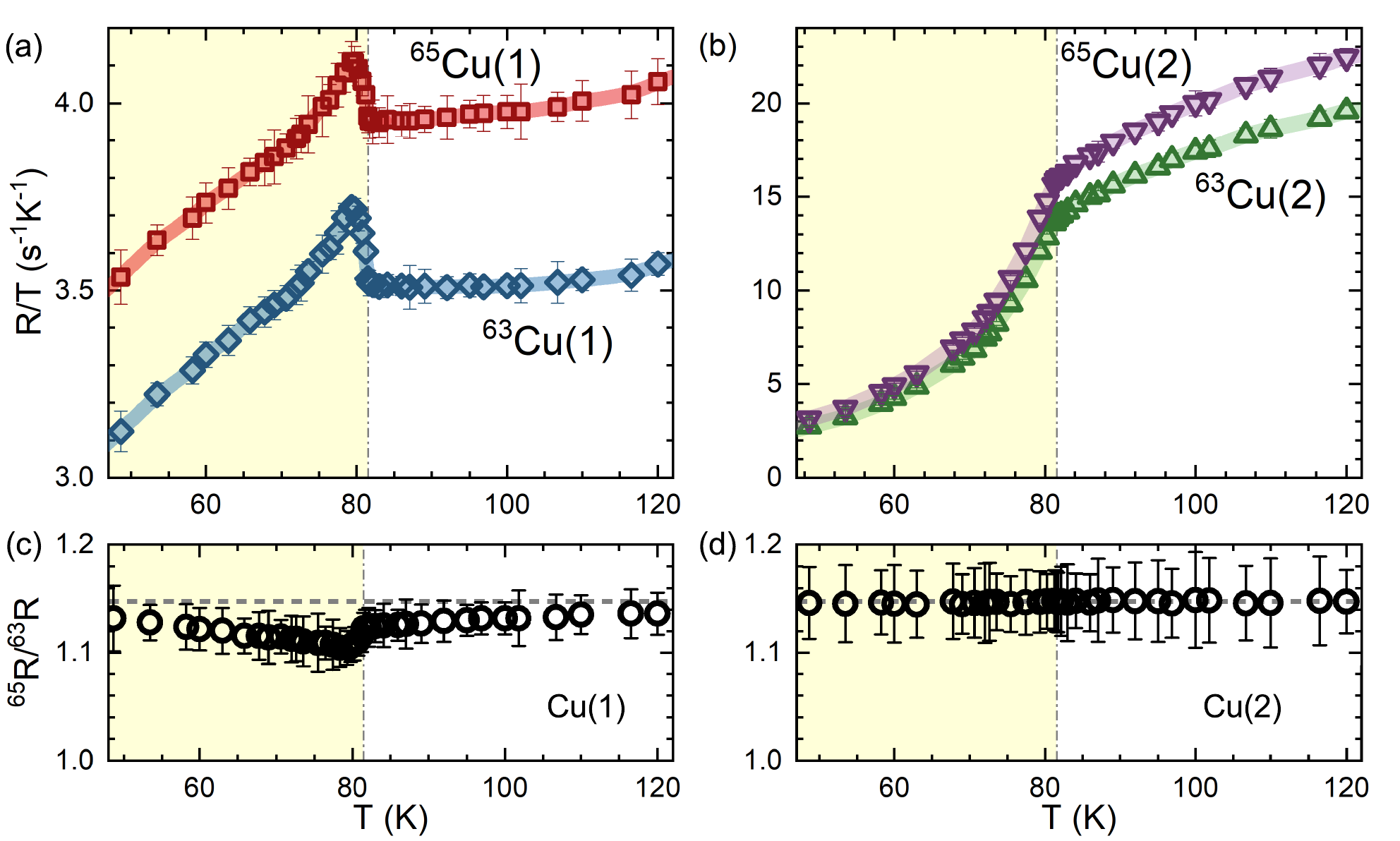}
\caption{Temperature dependence of spin-lattice relaxation rates and their ratio of YBa$_2$Cu$_4$O$_8$
around $T_{\rm c}$. The superconducting state is marked in yellow. Spin-lattice relaxation rates of \textsuperscript{65}Cu and \textsuperscript{63}Cu at the (a) Cu(1) site (cyan and red) and (b) Cu(2) site (purple and green), respectively. A coherence peak can be identified below $T_{\rm c}$ at the Cu(1) site. (c),(d) Spin-lattice relaxation rate ratios of \textsuperscript{65}Cu to \textsuperscript{63}Cu at Cu(1) and Cu(2) sites, respectively. The dashed lines mark the value of $(^{65} \gamma / ^{65} \gamma  )^2 = 1.1477$, where there is only magnetic relaxation. Deviation from the dashed line indicates quadrupole relaxation emerges below $T_{\rm c}$ at the Cu(1) site.}\label{fig:T1T}
\end{figure*}

Whether a coherence peak is present in $1/T_1^{\rm quad}$ motivated us to perform NQR on two isotopes $^{63}$Cu and $^{65}$Cu in oxygen-stoichiometric and underdoped YBa$_2$Cu$_4$O$_8$. This compound is an unconventional superconductor with a transition temperature $T_{\rm c} = 81.5$ K\cite{Karpinski1988}, as confirmed by the temperature-dependent magnetization measurement in Fig. \ref{fig:chi}. As previously mentioned, NQR is particularly sensitive to the local chemical environment and can distinguish between Cu atoms situated in distinct positions. YBa$_2$Cu$_4$O$_8$ contains two Cu sites, namely, the chain Cu(1) and the planar Cu(2) (depicted in the upper inset in Fig. \ref{fig:T1}), offering a unique avenue to probe $1/T_{1}^{\rm mag}$ and $1/T_{1}^{\rm quad}$ and explore the coherence peak. The Cu(1) and Cu(2) sites have different EFG strengths, resulting in different resonance frequencies as shown in the spectra inserted in Fig. \ref{fig:T1}. Additionally, the distinct EFG tensors of Cu(1) and Cu(2) influence their electric dynamic behavior, which is discussed later. Two isotopes \textsuperscript{63}Cu and \textsuperscript{65}Cu lead to two NQR resonance peaks at each site, for a total of four peaks at both sites, as shown in the lower inset in Fig. \ref{fig:T1} for the spectrum at $T_{\rm c}$.

The four NQR spin-lattice relaxation rates $R$ corresponding to the four peaks are measured and plotted in Fig. \ref{fig:T1}. For the planar Cu(2), the $R$ of \textsuperscript{63}Cu(2) and \textsuperscript{65}Cu(2) (green and purple lines, respectively) drop due to the reduction in DOS caused by the superconducting gap below $T_{\rm c}$, with noticeable kink behaviors observed at $T_{\rm c}$. The $R$ of Cu(2) above $T_{\rm c}$ keeps increasing and saturates around 200 K\cite{Raffa1998}. These data are consistent with previous reports in the whole temperature range\cite{Zimmermann1989, Machi1991} and similar to YBa$_2$Cu$_3$O$_7$\cite{Warren1990}. On the other hand, the $R$ of \textsuperscript{63}Cu(1) and \textsuperscript{65}Cu(1) (cyan and red lines, respectively) decrease with temperature decreasing above $T_{\rm c}$, resembling the behavior of a conventional metal. Notably, the $R$ of Cu(1) exhibits a slight increase just below $T_{\rm c}$, followed by a decrease at lower temperatures. These distinctive characteristics are highly unusual, suggesting a coherence peak akin to that observed in conventional superconductors.

Let us focus our attention on the superconducting transition and eliminate the linear temperature term in the metal by drawing
$R$/\emph{T} around $T_{\rm c}$ in Figs. \ref{fig:T1T}(a) and \ref{fig:T1T}(b). Both $R$/\emph{T} of \textsuperscript{63}Cu and \textsuperscript{65}Cu at the Cu(1) site exhibit a coherence peak just below $T_{\rm c}$, whereas $R$/\emph{T} of the Cu(2) site commences a quite steep descent below $T_{\rm c}$. The distinct temperature-dependence behaviors of Cu(1) and Cu(2) are attributed to their different EFG. $R$ can be expressed as a sum of magnetic relaxation rate $1/T_{1}^{\rm mag}$ and quadrupole relaxation rate $1/T_{1 }^{\rm quad}$, $R = 3/T_1^{{\text{mag}}} + 1/T_1^{{\text{quad}}}$ for \emph{I} = 3/2 NQR experiment\cite{Suter1998}. Theoretically, the nuclear spin Hamiltonian of the interaction between quadrupole moment \emph{Q} and EFG can be written as\cite{Slichter1990}
\begin{equation} \label{eq:vQ} 
\begin{aligned}
\mathcal{H}_{\rm Q} = \frac{h\nu_{\rm Q}}{6}\left[ (3I_{z}^{2}-I^{2})+\frac{\eta}{2} (I_{+}^{2}+I_{-}^{2})  \right]
\end{aligned}
\end{equation}
where $\nu_{\rm Q}$ is the quadrupole resonance frequency along the principal axis, $h\nu _{\rm Q}  = 3eQV_{zz} /2I(2I - 1)$. $\eta  = ( {V_{xx} - V_{yy} } )/V_{zz}$ is an asymmetry parameter of the EFG, where $V_{xx}, V_{yy}$, and $V_{zz}$ are the EFGs along the \emph{x}, \emph{y}, and \emph{z} directions, respectively. Only the terms with $I_+$ and $I_-$ can flip spins and contribute to $1/T_{1 }^{\rm quad}$, necessitating a sufficiently large $\eta$ to measure $1/T_{1 }^{\rm quad}$. The planar Cu(2) is isotropic with a negligible $\eta$ \cite{Zimmermann1989}, resulting in $1/T_1$ of Cu(2) being purely affected by magnetic relaxation and unable to detect EFG fluctuations\cite{Imai1988Cu2}. Therefore, although there are charge
fluctuations in the Cu-O plane detected by planar $^{17}$O with $\eta > 0.2$ \cite{Mangelschots1992}, they cannot be detected by Cu(2)\cite{Suter2000}. On the other hand, the chain Cu(1) is anisotropic with $\eta$ = 0.85, causing $1/T_1$ of Cu(1) to be affected by both magnetic and quadrupole relaxation. If the coherence peak is due to magnetic relaxation, it should be observed at both Cu(1) and Cu(2) sites. The presence of the coherence peak exclusively at the Cu(1) site suggests its origin from quadrupole relaxation.

$1/T_1^{\rm mag}$ and $1/T_1^{\rm quad}$ can be decomposed by utilizing two isotopes $^{63 , 65}$Cu, which possess the same spin \emph{I}=3/2 but different gyromagnetic ratios and quadrupole moments\cite{Imai1988Cu2}.
Using isotopes can avoid the influence from form factors and principle axes directions\cite{Goto1998}, since they occupy the same atom site.
The decomposition equations are $3/^{{\text{63}}} T_1 ^{{\text{mag}}}  = \left( {{}^{65}R - b \times ^{63}R} \right)/\left( {a - b} \right)
$ and $1/^{63}T_1^{{\text{quad}}}  = \left( {a \times {}^{63}R - ^{65}R} \right)/\left( {a - b} \right)
$, where the quotients $a = ( ^{65}\gamma /^{63}\gamma )^2 = 1.1477$ and $b = (^{65}Q/^{63}Q )^2 = 0.8562$ are known\cite{Raffa1999}.
Before decomposing, the ratio of $^{65}R/^{63}R$ is calculated to assess the weight of $1/T_1^{\rm mag}$ and $1/T_1^{\rm quad}$, shown in Figs. \ref{fig:T1T}(c) and \ref{fig:T1T}(d). $^{65}R/^{63}R$ at the Cu(2) site remains a constant value around the magnetic quotient $a=1.1477$ for various temperatures, indicating negligible $1/T_1^{\rm quad}$ and the presence of only $1/T_1^{\rm mag}$. Conversely, $^{65}R/^{63}R$ at the Cu(1) site deviates from $1.1477$ around $T_{\rm c}$, suggesting a detectable $1/T_1^{\rm quad}$ component.

\begin{figure}
\includegraphics[width=0.48\textwidth,clip]{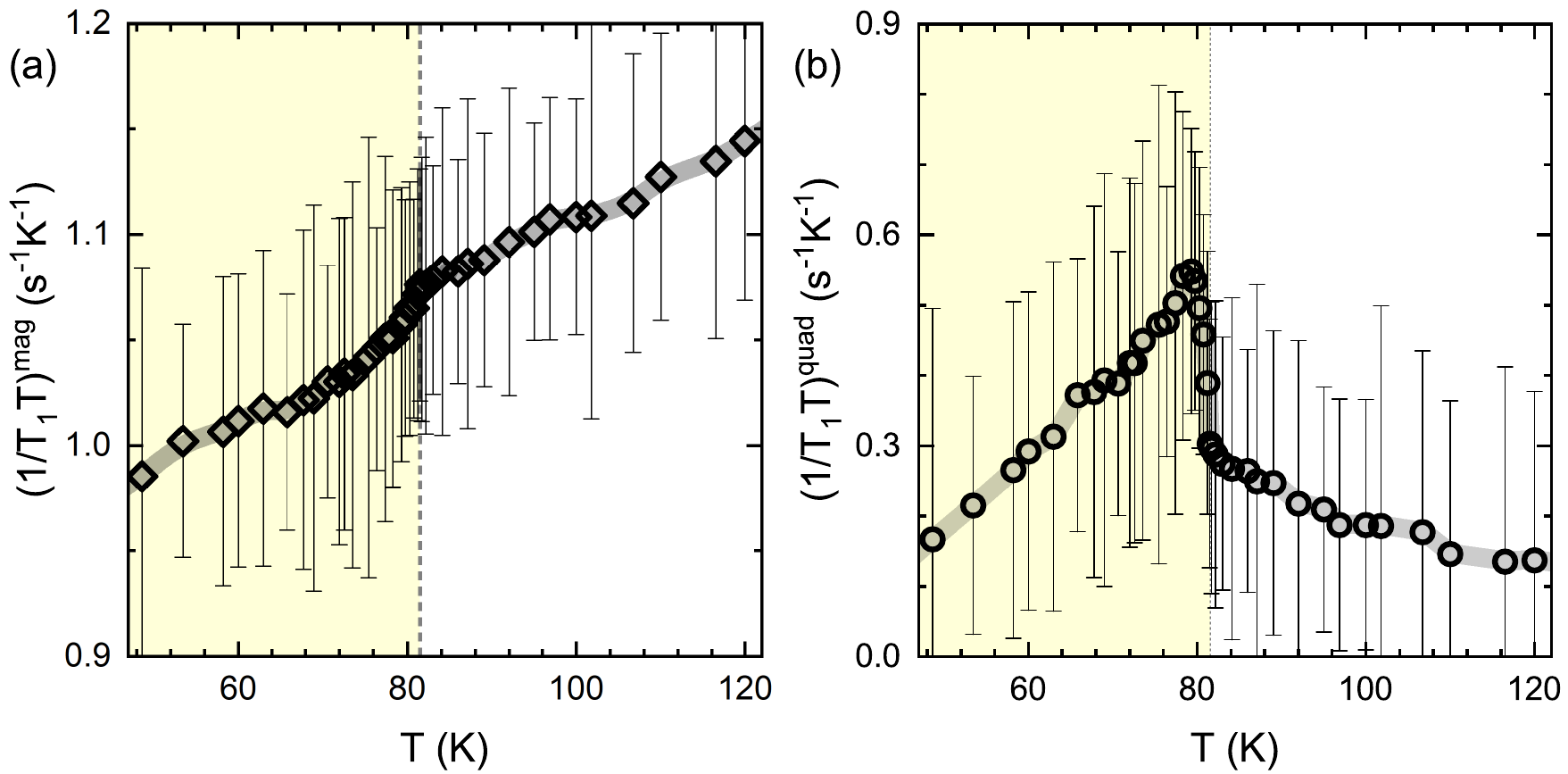}
\caption{The relaxation rate of (a) magnetic component $1/(T_{1}T)^{\rm mag}$ and (b) the quadrupole component $1/(T_{1 }T)^{\rm quad}$ at the Cu(1) site. The yellow background indicates the superconducting state. The coherence peak is located at the $1/(T_{1 }T)^{\rm quad}$ but is absent at the $1/(T_{1 }T)^{\rm mag}$.}\label{fig:T1MQ}
\end{figure}

The magnetic and quadrupole relaxations of Cu(1) are decomposed using the relations mentioned above, and the results are illustrated in Figs. \ref{fig:T1MQ}(a) and \ref{fig:T1MQ}(b), respectively. The $1/T_1^{\rm mag}$ depicted in Fig. \ref{fig:T1MQ}(a) drops below $T_{\rm c}$ due to the loss of DOS, aligning with the behavior of the $R/T$ at the Cu(2) site where solely magnetic relaxation occurs and $R/T = 3/( T_1 T )^{{\text{mag}}}$. In contrast, the $1/(T_1 T)^{\rm quad}$ displays a prominent coherence peak just below $T_{\rm c}$, in sharp contrast to conventional superconductors where the coherence peak is evident in the $1/(T_1 T)^{\rm mag}$ and absent in $1/(T_1 T)^{\rm quad}$. This discrepancy elucidates the absence of a coherence peak at the Cu(2) site, as it is unaffected by quadrupole relaxation. It is noteworthy that the $1/(T_1 T)^{\rm quad}$ is smaller than $1/(T_1 T)^{\rm mag}$, underscoring the higher sensitivity of most nuclei, such as Cu, to magnetic fluctuations in dipolar interaction than to EFG fluctuations in quadrupole interaction. Moreover, in the superconducting state, on lowering the temperature, the quadrupolar relaxation diminishes faster than the magnetic one\cite{Suter2000}. Consequently, the coherence peaks in Fig. \ref{fig:T1T} (a) have historically been inconspicuous and challenging to detect.

There is one thing we need to emphasize: The superfluid density of the system around $T_{\rm c}$ is dominated by the CuO\textsubscript{2} plane, since the bare transition temperature of the chain is much smaller than the plane owing to the strong fluctuations in one dimension \cite{Xiang1996,Serafin2010,Atkinson1995,Gagnon1997}. Therefore, although $1/T_{1 }^{\rm quad}$ is measured at the chains, the peak from $1/T_{1 }^{\rm quad}$ directly probes the charge dynamics of the CuO\textsubscript{2} plane.
To simplify our discussion and arrive at a qualitative understanding, we focus on only the CuO$_2$ plane and leave the general discussion to the appendix.
The newly observed $1/T_{1 }^{\rm quad}$ peak is proportional to a $q$-summed charge susceptibility $\chi _\rho  \left( {q,E} \right)$ of cuprate superconductivity, underscoring the significance of both long and short wavelengths to $1/T_{1 }^{\rm quad}$. Theoretically, $1/T_{1 }^{\rm quad}$ can be written as
\begin{widetext}
\begin{equation}\label{T1quad}
\frac{1}
{{T_1^{{\text{quad}}} }} \propto  - T\sum\limits_{kq} {\left( {1 - \frac{{\Delta _k \Delta _{k + q} }}
{{E_k ^2 }}} \right)F^2 \left( q \right)\frac{{\partial f\left( {E_k } \right)}}
{{\partial E_k }}\delta (E_k  - E_{k + q} )}
\end{equation}
\end{widetext}
where the $F( q )$ is the structure factor of the quadrupole interaction. A key aspect of this equation is the coherence factor $\left( {1 - {{\Delta _k \Delta _{k + q} } \mathord{\left/ {\vphantom {{\Delta _k \Delta _{k + q} } {E^2 }}} \right. \kern-\nulldelimiterspace} {E^2 }}} \right)$. For a \emph{d}-wave superconductor, a prominent momentum \emph{q} is around $q_0 = ( \pi ,\pi )$. This $q_0$ excitation, along with the corresponding coherence factor, has previously resulted in a distinct spin resonance peak in neutron scattering of cuprate superconductors\cite{Fong1995}. Similarly, this coherence factor around $q_0$ gives rise to a sign reversal $\Delta _k \Delta _{k + q_{\text{0}} }  < 0$, which cannot counterbalance the divergence from DOS in the superconducting state. Hence, this $q_0$ may lead to this unconventional coherence peak in $1/T_{1 }^{\rm quad}$ upon entering the superconducting transition.

\begin{figure*}
\includegraphics[width=0.75\textwidth,clip]{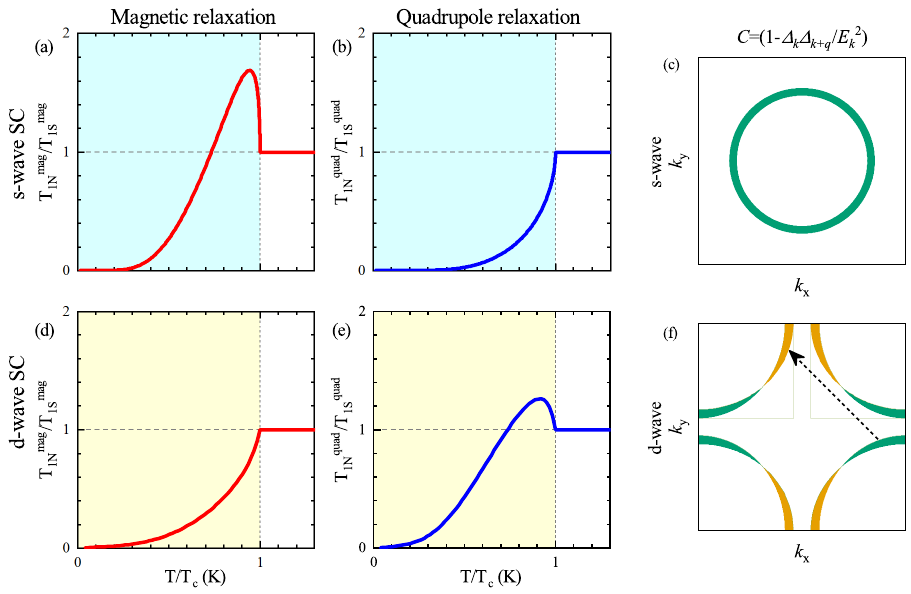}
\caption{Schematic diagrams of experimental measured magnetic and quadrupole relaxation rates of conventional and unconventional superconductors. The cyan and yellow backgrounds represent the conventional and unconventional superconducting states, respectively. (a) The magnetic relaxation rate shows a Hebel-Slichter coherence peak just below $T_{\rm c}$, while (b) the quadrupole relaxation rate drops rapidly below $T_{\rm c}$ in conventional superconductors. In unconventional superconductors, (d) the magnetic relaxation rate drops below $T_{\rm c}$, while (e) the quadrupole relaxation rate shows an unconventional coherence peak. (c) The Fermi surface of the $s$-wave superconducting gap, where the gap is isotropic. (f) The Fermi surface of the \emph{d}-wave superconducting gap, where the gap changes sign from positive (green) to negative (yellow). The arrow describes the scattering from positive to negative gap value, which leads to a sign change $\Delta _k \Delta _{k + q}  < 0$ in the quadrupole relaxation rate coherence factor $C = 1 - \Delta _k \Delta _{k + q} / E_k ^2 $.}\label{fig:sim}
\end{figure*}

\section{Discussion}
The unconventional coherence peak identified in the nuclear quadrupole relaxation rate of YBa$_2$Cu$_4$O$_8$ complements the relaxation rate pattern depicted in Fig. \ref{fig:sim}. In conventional superconductors, a Hebel-Slichter coherence peak is typically observed in the magnetic relaxation rate, as illustrated in Fig. \ref{fig:sim} (a), whereas the quadrupole relaxation rate drops below $T_{\rm c}$ due to coherence factors, as illustrated in Fig. \ref{fig:sim} (b). It reflects that the gap is isotropic without a sign reversal in conventional superconductors, as shown in Fig. \ref{fig:sim} (c). Conversely, unconventional superconducting gaps exhibit a sign reversal attributed to electron-electron correlation. In such cases, the magnetic relaxation rate does not display a coherence peak around $T_{\rm c}$ experimentally, and the quadrupole relaxation rate acquires a coherence peak, as illustrated in Figs. \ref{fig:sim}(d) and \ref{fig:sim}(e).
Historically, the absence of the Hebel-Slichter coherence peak in $1/T_{1}^{\rm mag}$ was widely discussed \cite{Imai1995, rigamonti_1998, ASAYAMA1996221,koyama_PhysRevB.39.2279,pines1990,pines1993, zoli1991, statt1990}. One of the prevailing views is that the sign-changing gap is the reason for this absence, which eliminates the coherence factor contribution \cite{pines1990,pines1993,koyama_PhysRevB.39.2279,Xiang2022}.
Meanwhile, it will lead to a phenomenon that the $1/T_{1}^{\rm quad}$ be enhanced by the coherence factors just below $T_{\rm c}$ due to the gap sign changes at various positions around Fermi surfaces, with scattering between different signs becoming dominant, as shown in Fig. \ref{fig:sim} (f).

The sharp contrasts observed in $1/T_{1}^{\rm mag}$ and $1/T_{1}^{\rm quad}$ from conventional superconductors provide a novel method to explore the unconventional nature of the pairing mechanism in unconventional superconductors. The spin-singlet Cooper pairs are widely believed to be formed above $T_{\rm c}$ owing to the small superfluid density in unconventional superconductors\cite{Emery1995}. The unconventional peak observed in $1/T_{1}^{\rm quad}$ leads to a hallmark of forming phase coherent unconventional superconducting condensate below $T_{\rm c}$. This peak can be used to diagnose unconventional (sign-changing gap) superconductivity with significant quadrupole relaxation.

On the other hand, measurement conditions for $1/T_{1}^{\rm quad}$ are quite stringent, necessitating substantial EFG fluctuations, significant $\eta$ and the presence of two or more isotopes. Cuprates have charge-density wave (CDW) fluctuations in the phase diagram\cite{Keimer2015, Wu2011}, which provide opportunity to detect $1/T_{1}^{\rm quad}$. Although CDW fluctuations contribute to $1/T_{1}^{\rm quad}$, the peak found here cannot be due to CDW which competes with superconductivity and CDW order appears only when superconductivity is killed. Some cuprates contain chain Cu with large $\eta$ and two isotopes \textsuperscript{63,65}Cu. Underdoped YBa$_2$Cu$_3$O$_{7-\delta}$ sharing similar electronic band structure with YBa$_2$Cu$_4$O$_{8}$ can also be utilized to identify the unconventional coherence peak\cite{Oguchi1990}. However, measuring $1/T_{1 }^{\rm quad}$ in other unconventional superconductors poses challenges, such as in iron arsenide where only one isotope of \textsuperscript{75}As is present\cite{LZ2011}. Theoretically, $1/T_{1}^{\rm quad}$ can be estimated by comparing the relaxation rate measured by NMR and NQR. However, the different principle axes and the external field make it hard to compare NMR and NQR directly\cite{Goto1998}. Another method is to compare the relaxation rate of satellite peaks with the central peak. Nevertheless, achieving this requires exceedingly precise data that surpass current experimental precision levels\cite{Suter1998}. As measurement precision improves, there may be opportunities to explore a wider range of systems. We hope our study will inspire extensive future experimental and theoretical investigations to elucidate whether the unconventional coherence peak is a universal characteristic of unconventional superconductivity and delve into its underlying mechanisms.

\begin{acknowledgments}
We thank Professor Chengtian Lin of Max-Planck-Institute für Festk\"{o}rperforschung for his help in sample synthesis. This work was supported by the National Key Research and Development Program of China (Grants No. 2022YFA1403903, No. 2022YFA1602800, and No. 2022YFA1403901), the National Natural Science Foundation of China (Grants No. 12134018, No. 12174428, No. 11888101, and No. 12488201), the Strategic Priority Research Program and Key Research Program of Frontier Sciences of the Chinese Academy of Sciences (Grant No. XDB33010100), the Chinese Academy of Sciences Project for Young Scientists in Basic Research (2022YSBR-048), the New Cornerstone Investigator Program, and the Synergetic Extreme Condition User Facility (SECUF).

\end{acknowledgments}

\appendix

\section{METHODS}

\subsection{Sample growth and characterization}
\begin{figure}
  \centering
  \includegraphics[width=0.45\textwidth,clip]{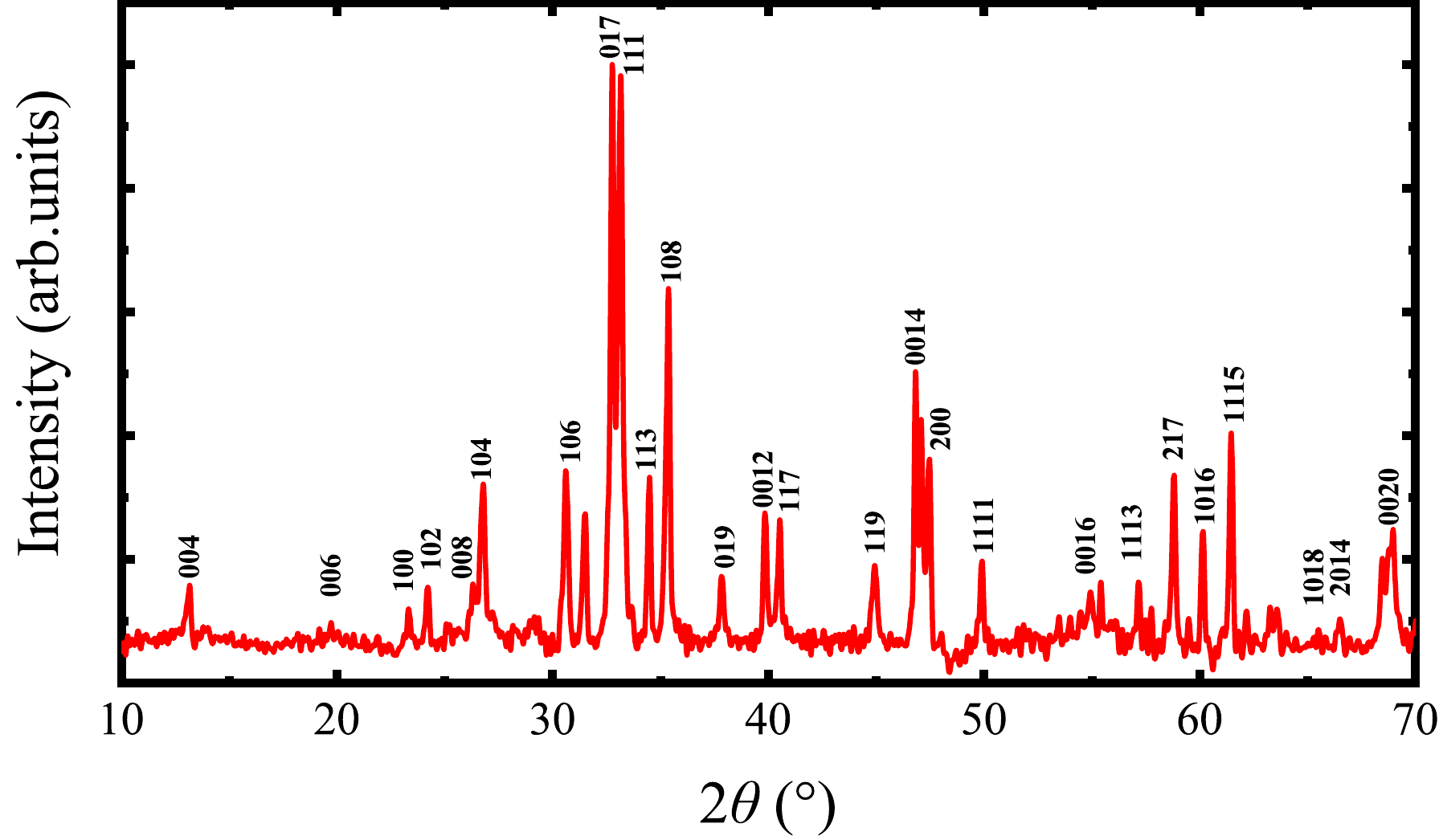}
  \caption{The x-ray diffraction pattern of YBa$_2$Cu$_4$O$_8$ at room temperature.}\label{fig:XRD}
\end{figure}

\begin{figure}
  \centering
  \includegraphics[width=0.35\textwidth,clip]{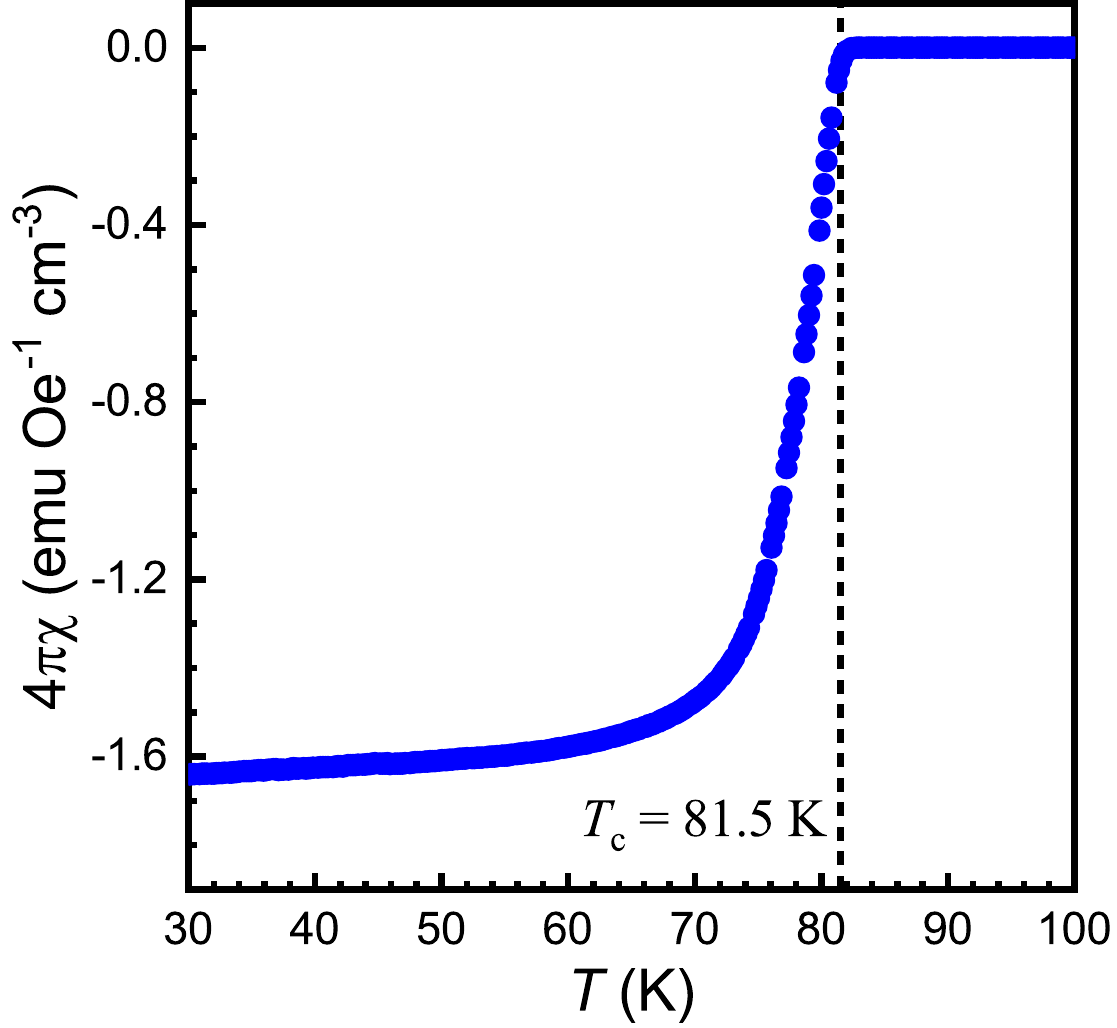}
  \caption{Temperature-dependent magnetic susceptibility of YBa$_2$Cu$_4$O$_8$ single crystal measured at 15 Oe. The data measured with zero field cooling show perfect diamagnetic. The dashed line indicates the superconducting transition temperature.}\label{fig:chi}
\end{figure}

YBa$_2$Cu$_4$O$_8$ single crystals are synthesized with YBa$_2$Cu$_3$O$_{7-\delta}$  powders and the same molar ratio CuO (99.9\%). The precursor YBa$_2$Cu$_3$O$_{7-\delta}$  is prepared by solid-state reaction. Stoichiometric proportions of Y$_2$O$_3$(99.99\%), BaCO$_3$(99.99\%), and CuO(99.9\%) are thoroughly mixed and grounded and then calcined in air at $860$ $^{\circ}$C for 24 h. The products are ground and then pressed into a pellet and calcined at 890 $^{\circ}$C for 48 h in Ar(95\%)-O2(5\%). The prepared YBa$_2$Cu$_3$O$_{7-\delta}$ powders and CuO are mixed uniformly and then put into the Al$_2$O$_3$ crucible with $50 \sim 70$ wt\% KOH as flux. After keeping at 700 $^{\circ}$C for 4 h, samples are cooled to 500 $^{\circ}$C at a speed of 8 $^{\circ}$C/h and fast cooled to room temperature finally. By soaking them in ethanol to remove the flux, we get small single crystals with a typical size of 0.1 mm. The x-ray diffraction pattern demonstrates the samples are YBa$_2$Cu$_4$O$_8$, as shown in Fig. \ref{fig:XRD}. The magnetic susceptibility measured with a magnetic property measurement system (MPMS-III) exhibits perfect diamagnetism, as shown in Fig. \ref{fig:chi}.

\subsection{NQR measurements}
\begin{figure}
  \centering
  \includegraphics[width=0.45\textwidth,clip]{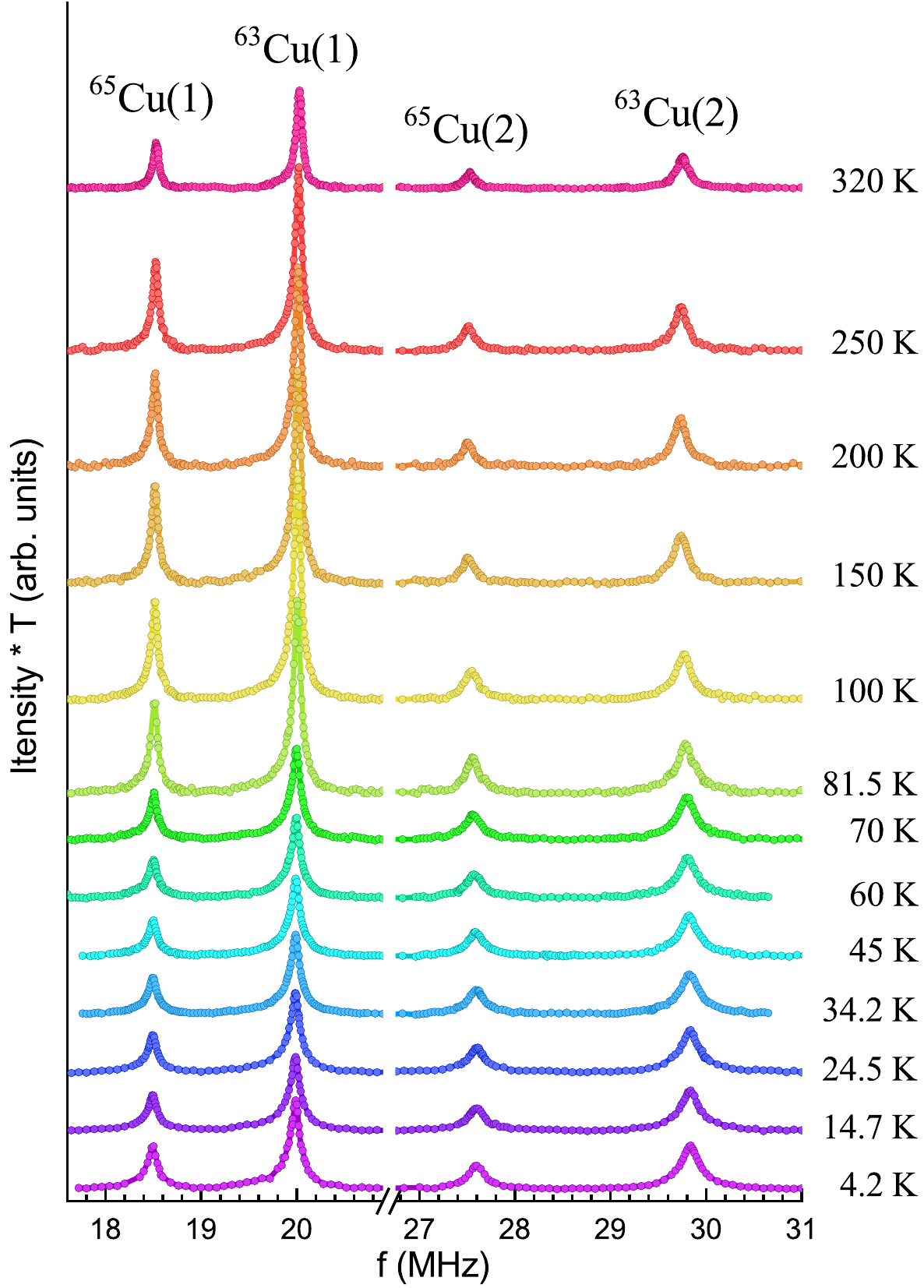}
  \caption{NQR spectra of YBa$_2$Cu$_4$O$_8$ at some typical temperatures. The curves are vertically offset for clarity. Four peaks are \textsuperscript{65,63}Cu of the Cu(1) site and Cu(2) site, respectively. The nature abundance of \textsuperscript{63}Cu and \textsuperscript{65}Cu are 69 \% and 31 \% respectively, which lead to a peak intensity of \textsuperscript{63}Cu about twice that of \textsuperscript{65}Cu.}\label{fig:Spectra}
\end{figure}

\begin{figure}
  \centering
  \includegraphics[width=0.4\textwidth,clip]{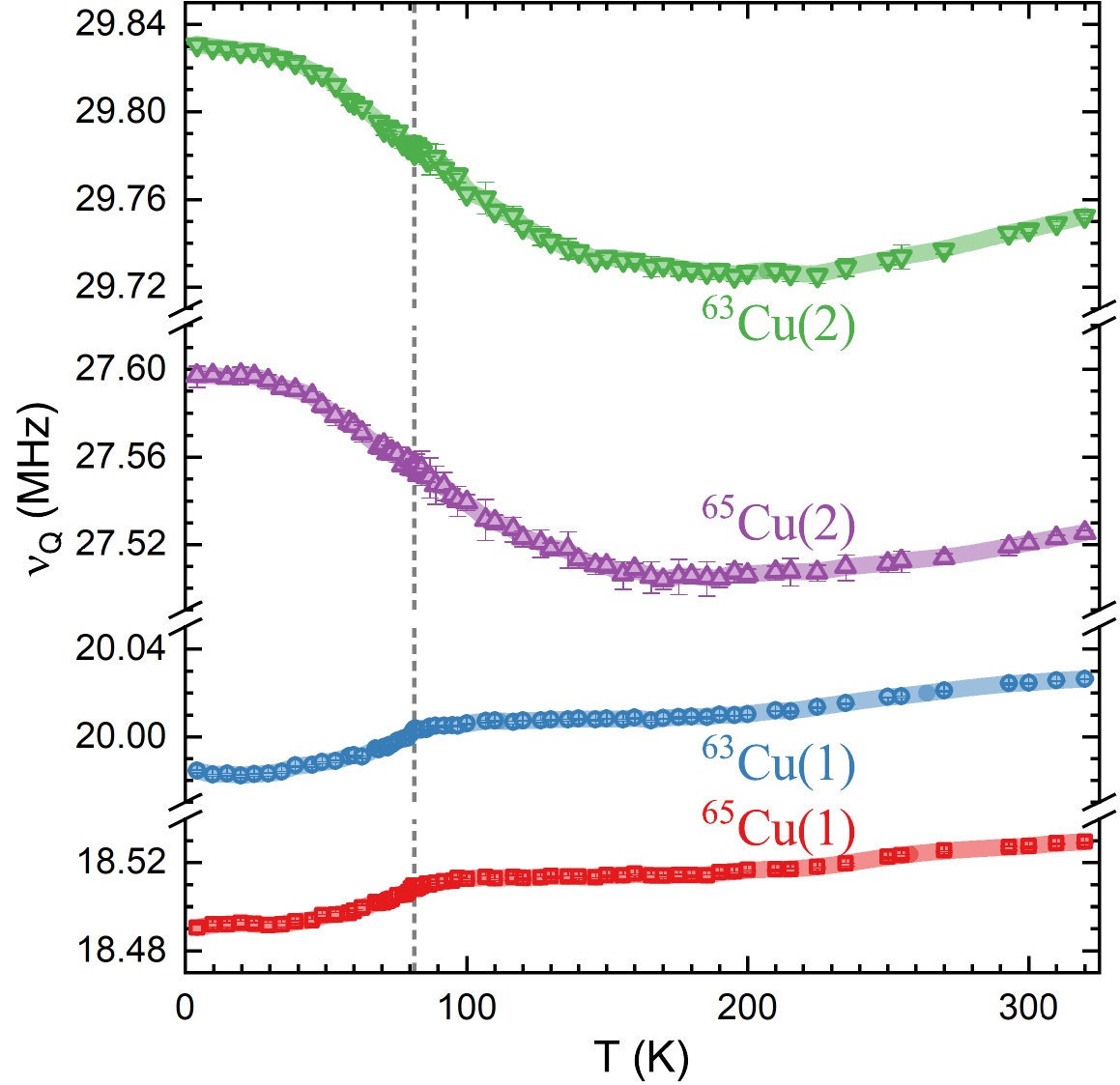}
  \caption{Temperature-dependent quadrupole resonance frequencies of \textsuperscript{63,65}Cu of the Cu(1) site and Cu(2) site. $T_{\rm c}$ is marked with a dashed line. The frequencies change little with temperature over the entire temperature range, even in the superconducting state. The spin-lattice relaxation rate at each temperature is measured at the corresponding frequencies.}\label{fig:vQ}
\end{figure}

\begin{figure}
  \centering
  \includegraphics[width=0.4\textwidth,clip]{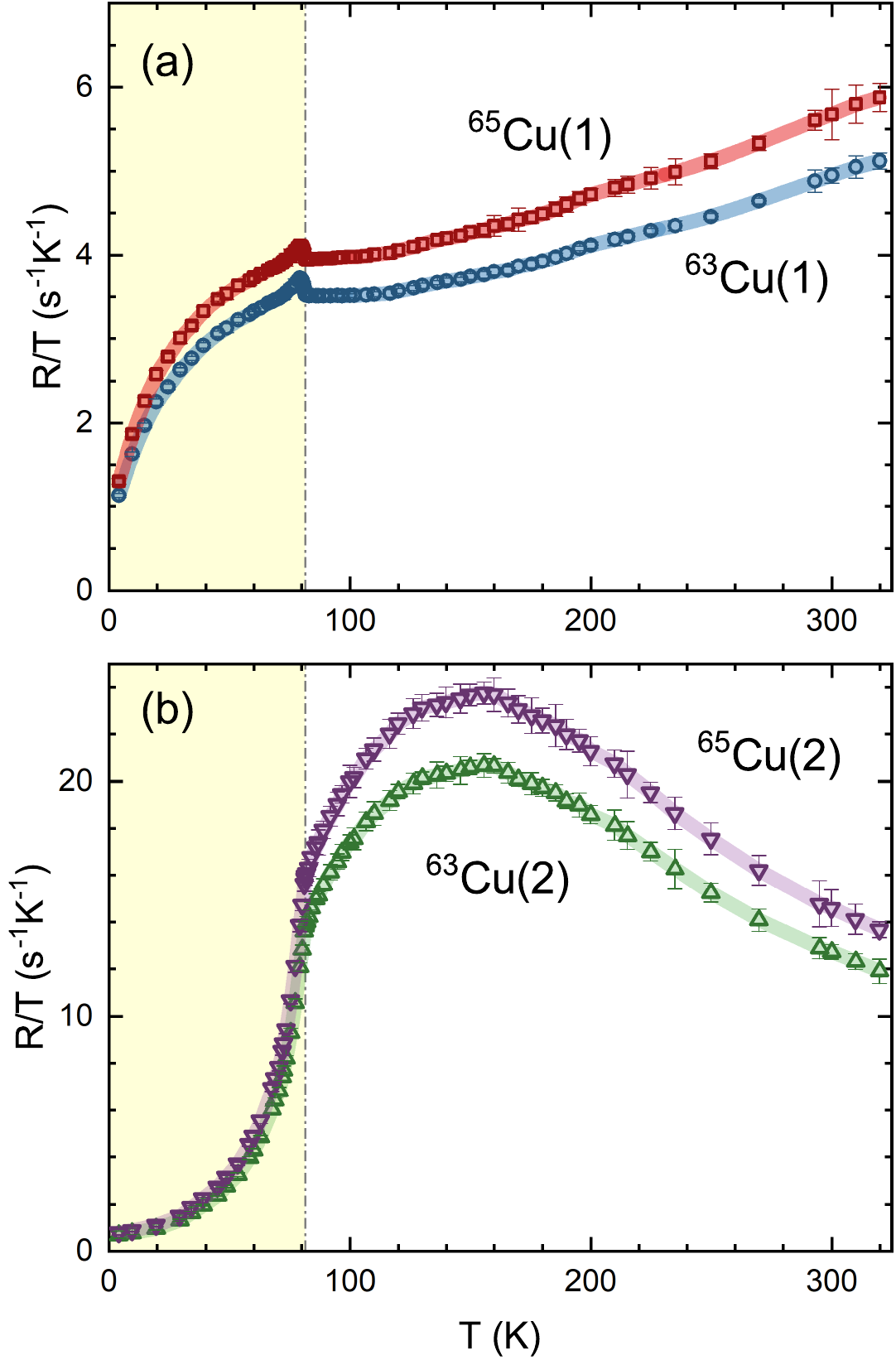}
  \caption{Temperature-dependent spin-lattice relaxation rates $R$/\emph{T} of \textsuperscript{65}Cu and \textsuperscript{63}Cu at the Cu(1) site and the Cu(2) site, respectively. The superconducting state is marked in yellow.}\label{fig:RT}
\end{figure}

The skin effect of the metallic state and the penetration depth of the superconducting state can shield the detection signal into the sample, so to achieve a large detectable volume we ground the samples and sieve with 300 mesh standard sieves to ensure a uniform powder particle size ($<50$ $\mu$m). NQR measurements are carried out using a commercial NMR spectrometer from Thamway Co. Ltd. The NQR spectra are acquired by integrating the intensity of spin echo at each frequency, as shown in Fig. \ref{fig:Spectra}. The quadrupole resonance frequencies $\nu_{\rm Q}$ are summarized in Fig. \ref{fig:vQ}.
We notice that the $\nu_{\rm Q}$ of Cu(1) shows a small kink at $T_{\rm c}$ and decreases below $T_{\rm c}$ in Fig. \ref{fig:vQ}.
It is natural to ask whether this kink can influence the spin-lattice relaxation rate $R$.
First of all, the change in $\nu_{\rm Q}$ is equal to the static EFG change crossing $T_{\rm c}$.
There are two possible origins for this static EFG change: (a) the electronic structure changes; (b) the coupling between electron and nuclear changes. Then, we can separate $\nu_Q$ change into two cases:
\begin{itemize}
    \item If the static EFG change comes from the electrons, the only electronic changes that occur at $T_{\rm c}$ are from the superconductivity. So, this case is equal to saying that the superconducting ordering induces both the $\nu_{\rm Q}$ change and $R$ peak crossing $T_{\rm c}$. And the $\nu_{\rm Q}$ change indicates that superconducting does influence the Cu(1) chain. In this case, $R$ is decided by superconducting transition but not $\nu_{\rm Q}$.
    \item If the static EFG change comes from the coupling, we can estimate this change to $R$. As discussed in the theoretical analysis subsection, the spin relaxation rate is proportional to the relaxation matrix $W_{mn}$ between a state $|m\rangle$ and $|n\rangle$ by $|\langle m|H_Q|n\rangle|^2$. Hence, the static EFG influence to the $R$ by the $(\nu_Q+\delta \nu_Q)^2/\nu_Q^2\approx1+2\delta \nu_Q/\nu_Q$. In other words, if we imagine the change of $R$ comes from a static coupling change instead of the electron fluctuations, the $R$ by the static EFG is changed by $2\delta \nu_Q/\nu_Q$.
\end{itemize}
We know that $\delta \nu_Q$ just below $T_c$ is less than $0.03\%$ in Fig. \ref{fig:vQ}. This means $0.03\%$ change of static EFG will change $R$ by $0.06\%$ in case (b). However, the $R$ in Fig. \ref{fig:T1T} (a) increases about $5\%$ just below $T_{\rm c}$ which is much larger than this small change. Moreover, $\nu_{\rm Q}$ decreases monotonically with decreasing temperature below $T_{\rm c}$, while $R$ peaks just below $T_{\rm c}$ and decreases at lower temperatures. So the peak of $R$ cannot come from $\nu_{\rm Q}$ change. The change of $R$ reflects the change of DOS and the appearance of the coherence factor in the superconducting state.

Spin-lattice relaxation rate $R$ for every nucleus is measured using a comb-shaped-pulse recovery method with the recovery function $M\left( t \right) = M(\infty)-A\exp ( { - Rt} )$, where error bars are the standard error of the least square fit. $M(t)$ is the nuclear magnetization at time \emph{t} after saturation pulses.  $M(\infty)$ is the value of $M(t)$ in an equilibrium state and $A=M(\infty)-M(0)$, where $M(0)$ is the initial value of $M(t)$ after the saturation pulses. Both $M(\infty)$ and $A$ are fitting parameters. $R/T$ in the full temperature zone is shown in Fig. \ref{fig:RT}. $R$ can be written as a sum of two contributions of $1/T_{1}^{\rm mag}$ and
$1/T_{1 }^{\rm quad}$, $^{63,65}R = 3/^{63,65}T_1^{{\text{mag}}} + 1/^{63,65}T_1^{{\text{quad}}}$ for \textsuperscript{63,65}Cu NQR experiment. $1/T_{1}^{\rm mag}$ is proportional to the square of the gyromagnetic ratio, ${\text{1/}}T_1^{{\text{mag}}}  \propto \gamma ^2$, so $a \equiv (^{63} T_1^{{\text{mag}}} /^{65} T_1^{{\text{mag}}} ) = (^{65} \gamma /^{63} \gamma )^2  = 1.1477$. $1/T_{1 }^{\rm quad}$ is proportional to the square of the quadruple moment, ${1 \mathord{\left/ {\vphantom {1 {T_1^{{\text{quad}}} }}} \right. \kern-\nulldelimiterspace} {T_1^{{\text{quad}}} }} \propto Q^2$, so $b \equiv (^{{\text{63}}} T_1^{{\text{quad}}} /^{65} T_1^{{\text{quad}}} ) = (^{65} Q/^{63} Q)^2  = 0.8562$. Based on these relationships,
$1/T_{1}^{\rm mag}$ and $1/T_{1 }^{\rm quad}$ can be distinguished.

We want to add a note here. The EFG principal axis of Cu(1) is along the $a$ axis and the principal axis of Cu(2) is along the $c$ axis\cite{Zimmermann1989}. In a conventional superconductor, the Hebel-Slichter coherence peak appears in all directions. The coherence peak reflects the divergence of DOS and coherence between electrons. Whether there is a coherence peak does not depend on the direction of the principal axes of EFG. However, if one wants to separate $1/T_1^{\rm mag}$ and $1/T_1^{\rm quad}$ by comparing NMR and NQR, or by comparing atoms at different sites, the principal axis is important. $1/T_1^{\rm mag}$ is determined by the fluctuations perpendicular to the applied magnetic field and $1/T_1^{\rm quad}$ is determined by the fluctuations perpendicular to the principal axes when $\eta=0$. The mismatch between the applied magnetic field and the principle axes leads to different $1/T_1$ values of NMR from NQR. When $\eta>0$, even if the applied field is along the principal axes, NMR and NQR cannot get the same $1/T_1$ value\cite{Goto1998}. Moreover, the principal axes of Cu(1) and Cu(2) are different, so their $1/T_1$ from NQR cannot be compared directly\cite{Zimmermann1989}. We use the method of comparing $1/T_1$ of $^{63}$Cu and $^{65}$Cu at the same site, which have the same EFG and form factor, to avoid this problem.

\subsection{Theoretical analysis}
\begin{figure}
  \centering
  \includegraphics[width=0.4\textwidth,clip]{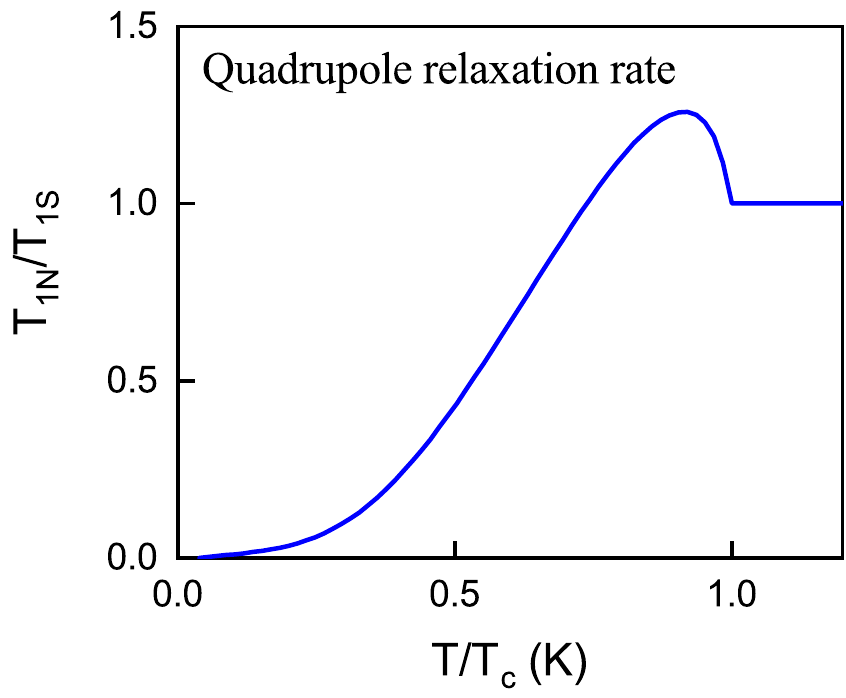}
  \caption{The theoretical simulation for the quadrupole relaxation rate in \emph{d}-wave superconductors. Here, we limit the summation around $\left( {\pi ,\pi } \right)$.}\label{fig:T1d}
\end{figure}

The Hamiltonian of quadrupole interaction can be compactly written as
\begin{equation}
    H_Q  = eQ\int {{\text{d}}rV_{\mu \nu } } \left( r \right)I_{\mu \nu } \left( r \right).
\end{equation}
The EFG tensor $V_{\mu\nu}$ can be expressed as $V_{\mu \nu } (r) = \int {{\text{d}}r'\rho (r')T_{\mu \nu } } \left( {r,r'} \right)$ where $T_{\mu \nu } \left( {r,r'} \right)$ is the spatial function linking the electron density $\rho(r')$ to nucleus quadrupole $I_{\mu\nu}(r)$. Thus, the charge fluctuation in NQR is mainly determined by the density fluctuation of electrons and quadrupole of the nuclear. The Fourier transformed Hamiltonian in the lattice can be expressed as $H_Q  = e\sum\limits_q {F(q)} \rho (q)A(- q)$, where $F(q)$ is the structure factor determined by $T_{\mu \nu } \left( {r,r'} \right)$ and \emph{A} contains the quadrupole moment and spin of nuclear. The relaxation of the nuclear spin and lattice toward the thermodynamic equilibrium can be described by the master equation $\frac{{{\text{d}}P(t)}} {{dt}} = W_{mn} \left[ {P(t) - P(0)} \right]$, where $P(t)$ is the population vector of the different energy levels \cite{Suter1998}. The relaxation matrix $W_{mn}$ is given by $W_{mn}  \propto T\sum\limits_q {F^2 (q)} \mathop {\lim }\limits_{\omega  \to 0} \frac{{\operatorname{Im} \chi _{\rho \rho } (q,\omega )}}{\omega }$ through second-order perturbation theory. Here $\chi _{\rho \rho } (q,\omega )$ is the density-density correlation function. For superconductors, $\chi _{\rho \rho } (q,\omega )$ can be calculated by the Green's function as
\begin{equation}
\begin{aligned}
    \chi _{\rho \rho} (q,\text{i}\omega_n) &=& F^2 (q)\frac{1}{\beta }\sum_{k,\text{i}\upsilon_n} \text{Tr}\left[ G(k,\text{i}\upsilon_n )\tau _3 \right. \\
    && \left.G(k + q,\text{i}\upsilon_n  + \text{i}\omega_n )\tau _3  \right],
    \end{aligned}
\end{equation}
where $\tau _3 $ is the Pauli matrix and $G(k,{\text{i}}\upsilon _n )$ is the Matsubara Green's function for SC. Then we can get
\begin{equation}
    \frac{1}{T_1^{\text{quad}}} \propto - T\sum_{kq} {\left( {1 - \frac{{\Delta _k \Delta _{k + q} }}{{E_k ^2 }}} \right)F^2 \left( q \right)\frac{{\partial f\left( {E_k } \right)}}{{\partial E_k}}\delta (E_k - E_{k + q} )}.
\end{equation}
As we discuss in the main text, the coherence peak is from the CuO$_2$ plane. By focusing on the plane, Fig. \ref{fig:T1d} shows the quadrupole relaxation rate in \emph{d}-wave superconductors calculated by summing the $q$ points near $(\pi ,\pi)$.

\begin{figure}
  \centering
  \includegraphics[width=0.5\textwidth,clip]{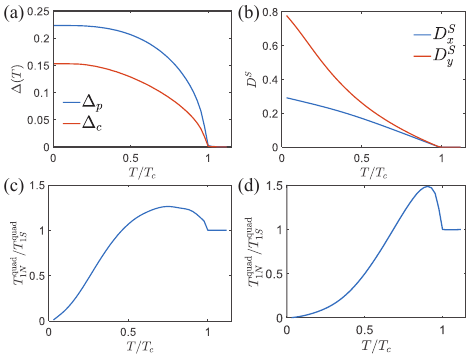}
  \caption{(a) The amplitude of order parameters on CuO chain $\Delta_c$ and plane $\Delta_p$ through the self-consistent calculation. (b) The superfluid density along $x$ and $y$ directions. (c) The quadrupole relaxation rate on the CuO chain [Cu(1) site]. (d) The quadrupole relaxation rate on the CuO plane [Cu(2) site]. The parameters in calculation are set as $t_{pNN}=0.36$, $t_{pNNN}=0.12$, $t_{cNN}=0.45$, $t_{cp}=0.014$, $t_{cpNN}=0.04$, $\varepsilon_c=0.26$, $\varepsilon_p=0.22$, $\lambda_1=\lambda_2=0.8$ and $\lambda_J=0.15$.}\label{fig-chain}
\end{figure}

The above analysis provides a qualitative understanding of the coherence peak in $d$-wave superconductors. On the other hand, since NQR relaxation rate can be observed only on the Cu(1) sites rather than on the Cu(2) sites due to a nonzero $\eta$ on CuO chains, we need consider a multiband model including both CuO planes and CuO chains to more convincingly describe the experimental phenomena.
The Hamiltonian of the two-band model can be written as
\begin{equation}
\begin{aligned}
 H_0&=&\sum_{nk} \varepsilon_{n k} c_{n k \sigma}^{\dagger} c_{n k \sigma}+\varepsilon_{\perp k}\left(c_{1 k \sigma}^{\dagger} c_{2 k \sigma}+\text { H.c. }\right) \\
 && + \sum_{nk} \Delta_n\phi_{nk} \left(c_{nk\uparrow}^\dagger c_{n-k\downarrow}^\dagger +\text { H.c. }\right).
\end{aligned}
\end{equation}
The dispersions of the CuO chain ($n=1$) and CuO plane ($n=2$) are $\varepsilon_{1k}=-2 t_{cNN}\cos k_x+\varepsilon_c$ and $\varepsilon_{2k}=-2 t_{pNN}\left(\cos k_x+\cos k_y\right)+4t_{pNNN}\cos{k_x}\cos{k_y}+\varepsilon_p$ respectively. The tunneling term between the two layers has the form $\varepsilon_{\perp k}=t_{cp}+2t_{cpNN}(\cos{k_x}-\cos{k_y})$. All the parameters in the tight-binding model can be obtained from the density-functional theory calculations.
$\Delta_n\phi_{nk}$ is the mean-field order parameter of the chain and plane bands.
To our knowledge, how the chain becomes superconducting remains controversial. Since the failure of considering only the single-particle tunneling model has been highlighted in Ref. \cite{Xiang1996}, we take the pair tunneling interaction, specifically Josephson coupling.
The singlet pair tunneling term has the form
\begin{equation}
    H_I=-\lambda_J \sum_{kk'}\left(\phi_{1k}\phi_{2k'}c_{1k\uparrow}^\dagger c_{1-k\downarrow}^\dagger c_{2-k'\downarrow} c_{2k'\uparrow}+\text { H.c. }\right),
\end{equation}
where $\lambda_J$ is the Josephson coupling strength. Because of symmetry requirements, Josephson coupling imposes the same pairing symmetry on both the CuO chains and planes. Thus, we set $\phi_{1k}=\phi_{2k}=\cos{k_x}-\cos{k_y}$. Then $\Delta_n$ can be determined by self-consistent equations
\begin{equation}
    \begin{aligned}
        \Delta_1 &=& \frac{\lambda_1}{V} \sum_k \phi_{1 k}\left\langle c_{1 k \uparrow} c_{1-k \downarrow}\right\rangle + \frac{\lambda_J}{V}\sum_k \phi_{2 k}\left\langle c_{2 k \uparrow} c_{2-k \downarrow}\right\rangle, \\
        \Delta_2 &=& \frac{\lambda_J}{V} \sum_k \phi_{1 k}\left\langle c_{1 k \uparrow} c_{1-k \downarrow}\right\rangle + \frac{\lambda_2}{V}\sum_k \phi_{2 k}\left\langle c_{2 k \uparrow} c_{2-k \downarrow}\right\rangle,
    \end{aligned}
\end{equation}
where $\lambda_n$ are strengths of intralayer pairing potentials. Figure \ref{fig-chain}(a) shows the amplitude of order parameters on the CuO chain and plane. Because of Josephson coupling, the order parameter from the CuO planes penetrates into the chains. As a result, the phase coherence properties of the planes can also be detected in the chains. Note that if $\lambda_J=0$, the self-consistent calculation will yield a very small value for $\Delta_c$. Furthermore, we calculate the superfluid density along $x$ and $y$ directions shown in Fig.\ref{fig-chain}(b). The quadrupole relaxation rate projected on the chain and plane bands can be calculated using this two-band model. Figure \ref{fig-chain}(c) and \ref{fig-chain}(d) show the results. The quadrupole relaxation rate on the CuO chain qualitatively captures the main features in the experiment: The coherence peak occurs just below $T_c$. We emphasize that the conclusions do not depend on the specific form of interlayer coupling, as we also achieve the similar results using the single-particle tunneling model.


\begin{thebibliography}{39}%
\makeatletter
\providecommand \@ifxundefined [1]{%
 \@ifx{#1\undefined}
}%
\providecommand \@ifnum [1]{%
 \ifnum #1\expandafter \@firstoftwo
 \else \expandafter \@secondoftwo
 \fi
}%
\providecommand \@ifx [1]{%
 \ifx #1\expandafter \@firstoftwo
 \else \expandafter \@secondoftwo
 \fi
}%
\providecommand \natexlab [1]{#1}%
\providecommand \enquote  [1]{``#1''}%
\providecommand \bibnamefont  [1]{#1}%
\providecommand \bibfnamefont [1]{#1}%
\providecommand \citenamefont [1]{#1}%
\providecommand \href@noop [0]{\@secondoftwo}%
\providecommand \href [0]{\begingroup \@sanitize@url \@href}%
\providecommand \@href[1]{\@@startlink{#1}\@@href}%
\providecommand \@@href[1]{\endgroup#1\@@endlink}%
\providecommand \@sanitize@url [0]{\catcode `\\12\catcode `\$12\catcode
  `\&12\catcode `\#12\catcode `\^12\catcode `\_12\catcode `\%12\relax}%
\providecommand \@@startlink[1]{}%
\providecommand \@@endlink[0]{}%
\providecommand \url  [0]{\begingroup\@sanitize@url \@url }%
\providecommand \@url [1]{\endgroup\@href {#1}{\urlprefix }}%
\providecommand \urlprefix  [0]{URL }%
\providecommand \Eprint [0]{\href }%
\providecommand \doibase [0]{https://doi.org/}%
\providecommand \selectlanguage [0]{\@gobble}%
\providecommand \bibinfo  [0]{\@secondoftwo}%
\providecommand \bibfield  [0]{\@secondoftwo}%
\providecommand \translation [1]{[#1]}%
\providecommand \BibitemOpen [0]{}%
\providecommand \bibitemStop [0]{}%
\providecommand \bibitemNoStop [0]{.\EOS\space}%
\providecommand \EOS [0]{\spacefactor3000\relax}%
\providecommand \BibitemShut  [1]{\csname bibitem#1\endcsname}%
\let\auto@bib@innerbib\@empty
\bibitem [{\citenamefont {Taillefer}(2010)}]{Taillefer2012}%
  \BibitemOpen
  \bibfield  {author} {\bibinfo {author} {\bibfnamefont {L.}~\bibnamefont
  {Taillefer}},\ }\bibinfo {title} {Scattering and pairing in cuprate
  superconductors},\ in\ \href
  {https://doi.org/10.1146/annurev-conmatphys-070909-104117} {\emph {\bibinfo
  {booktitle} {Annual Review of Condensed Matter Physics, Vol 1}}},\ \bibinfo
  {series} {Annual Review of Condensed Matter Physics}, Vol.~\bibinfo {volume}
  {1},\ \bibinfo {editor} {edited by\ \bibinfo {editor} {\bibfnamefont {J.~S.}\
  \bibnamefont {Langer}}}\ (\bibinfo {year} {2010})\ pp.\ \bibinfo {pages}
  {51--70}\BibitemShut {NoStop}%
\bibitem [{\citenamefont {Imai}\ \emph
  {et~al.}(1988{\natexlab{a}})\citenamefont {Imai}, \citenamefont {Shimizu},
  \citenamefont {Tsuda}, \citenamefont {Yasuoka}, \citenamefont {Takabatake},
  \citenamefont {Nakazawa},\ and\ \citenamefont {Ishikawa}}]{Imai1988Cu2}%
  \BibitemOpen
  \bibfield  {author} {\bibinfo {author} {\bibfnamefont {T.}~\bibnamefont
  {Imai}}, \bibinfo {author} {\bibfnamefont {T.}~\bibnamefont {Shimizu}},
  \bibinfo {author} {\bibfnamefont {T.}~\bibnamefont {Tsuda}}, \bibinfo
  {author} {\bibfnamefont {H.}~\bibnamefont {Yasuoka}}, \bibinfo {author}
  {\bibfnamefont {T.}~\bibnamefont {Takabatake}}, \bibinfo {author}
  {\bibfnamefont {Y.}~\bibnamefont {Nakazawa}},\ and\ \bibinfo {author}
  {\bibfnamefont {M.}~\bibnamefont {Ishikawa}},\ }\bibfield  {title} {\bibinfo
  {title} {{Nuclear Spin-Lattice Relaxation of $^{63,65}$Cu at the Cu(2) Sites
  of the High $T_{\rm c}$ Superconductor YBa$_2$Cu$_3$O$_{7-\delta}$}},\ }\href
  {https://doi.org/10.1143/JPSJ.57.1771} {\bibfield  {journal} {\bibinfo
  {journal} {Journal of the Physical Society of Japan}\ }\textbf {\bibinfo
  {volume} {57}},\ \bibinfo {pages} {1771} (\bibinfo {year}
  {1988}{\natexlab{a}})}\BibitemShut {NoStop}%
\bibitem [{\citenamefont {Imai}\ \emph
  {et~al.}(1988{\natexlab{b}})\citenamefont {Imai}, \citenamefont {Shimizu},
  \citenamefont {Yasuoka}, \citenamefont {Ueda},\ and\ \citenamefont
  {Kosuge}}]{Imai1988Cu63}%
  \BibitemOpen
  \bibfield  {author} {\bibinfo {author} {\bibfnamefont {T.}~\bibnamefont
  {Imai}}, \bibinfo {author} {\bibfnamefont {T.}~\bibnamefont {Shimizu}},
  \bibinfo {author} {\bibfnamefont {H.}~\bibnamefont {Yasuoka}}, \bibinfo
  {author} {\bibfnamefont {Y.}~\bibnamefont {Ueda}},\ and\ \bibinfo {author}
  {\bibfnamefont {K.}~\bibnamefont {Kosuge}},\ }\bibfield  {title} {\bibinfo
  {title} {{Anomalous Temperature Dependence of Cu Nuclear Spin-Lattice
  Relaxation in YBa$_2$Cu$_3$O$_{6.91}$}},\ }\href
  {https://doi.org/10.1143/JPSJ.57.2280} {\bibfield  {journal} {\bibinfo
  {journal} {Journal of the Physical Society of Japan}\ }\textbf {\bibinfo
  {volume} {57}},\ \bibinfo {pages} {2280} (\bibinfo {year}
  {1988}{\natexlab{b}})}\BibitemShut {NoStop}%
\bibitem [{\citenamefont {Zimmermann}\ \emph {et~al.}(1989)\citenamefont
  {Zimmermann}, \citenamefont {Mali}, \citenamefont {Brinkmann}, \citenamefont
  {Karpinski}, \citenamefont {Kaldis},\ and\ \citenamefont
  {Rusiecki}}]{Zimmermann1989}%
  \BibitemOpen
  \bibfield  {author} {\bibinfo {author} {\bibfnamefont {H.}~\bibnamefont
  {Zimmermann}}, \bibinfo {author} {\bibfnamefont {M.}~\bibnamefont {Mali}},
  \bibinfo {author} {\bibfnamefont {D.}~\bibnamefont {Brinkmann}}, \bibinfo
  {author} {\bibfnamefont {J.}~\bibnamefont {Karpinski}}, \bibinfo {author}
  {\bibfnamefont {E.}~\bibnamefont {Kaldis}},\ and\ \bibinfo {author}
  {\bibfnamefont {S.}~\bibnamefont {Rusiecki}},\ }\bibfield  {title} {\bibinfo
  {title} {{Copper NQR and NMR in the superconductor YBa$_2$Cu$_4$O$_{8+x}$}},\
  }\href {https://doi.org/https://doi.org/10.1016/0921-4534(89)91304-X}
  {\bibfield  {journal} {\bibinfo  {journal} {Physica C: Superconductivity}\
  }\textbf {\bibinfo {volume} {159}},\ \bibinfo {pages} {681} (\bibinfo {year}
  {1989})}\BibitemShut {NoStop}%
\bibitem [{\citenamefont {Warren}\ and\ \citenamefont
  {Walstedt}(1990)}]{Warren1990}%
  \BibitemOpen
  \bibfield  {author} {\bibinfo {author} {\bibfnamefont {W.~W.}\ \bibnamefont
  {Warren}}\ and\ \bibinfo {author} {\bibfnamefont {R.~E.}\ \bibnamefont
  {Walstedt}},\ }\bibfield  {title} {\bibinfo {title} {{NQR and NMR Studies of
  Spin Dynamics in High $T_c$ Superconducting Cuprates}},\ }\href
  {https://doi.org/doi:10.1515/zna-1990-3-429} {\bibfield  {journal} {\bibinfo
  {journal} {Zeitschrift für Naturforschung A}\ }\textbf {\bibinfo {volume}
  {45}},\ \bibinfo {pages} {385} (\bibinfo {year} {1990})}\BibitemShut
  {NoStop}%
\bibitem [{\citenamefont {Machi}\ \emph {et~al.}(1991)\citenamefont {Machi},
  \citenamefont {Tomeno}, \citenamefont {Miyatake}, \citenamefont {Koshizuka},
  \citenamefont {Tanaka}, \citenamefont {Imai},\ and\ \citenamefont
  {Yasuoka}}]{Machi1991}%
  \BibitemOpen
  \bibfield  {author} {\bibinfo {author} {\bibfnamefont {T.}~\bibnamefont
  {Machi}}, \bibinfo {author} {\bibfnamefont {I.}~\bibnamefont {Tomeno}},
  \bibinfo {author} {\bibfnamefont {T.}~\bibnamefont {Miyatake}}, \bibinfo
  {author} {\bibfnamefont {N.}~\bibnamefont {Koshizuka}}, \bibinfo {author}
  {\bibfnamefont {S.}~\bibnamefont {Tanaka}}, \bibinfo {author} {\bibfnamefont
  {T.}~\bibnamefont {Imai}},\ and\ \bibinfo {author} {\bibfnamefont
  {H.}~\bibnamefont {Yasuoka}},\ }\bibfield  {title} {\bibinfo {title}
  {{Nuclear spin-lattice relaxation and Knight shift in YBa$_2$Cu$_4$O$_8$}},\
  }\href {https://doi.org/https://doi.org/10.1016/0921-4534(91)90790-6}
  {\bibfield  {journal} {\bibinfo  {journal} {Physica C: Superconductivity}\
  }\textbf {\bibinfo {volume} {173}},\ \bibinfo {pages} {32} (\bibinfo {year}
  {1991})}\BibitemShut {NoStop}%
\bibitem [{\citenamefont {Asayama}\ \emph {et~al.}(1996)\citenamefont
  {Asayama}, \citenamefont {Kitaoka}, \citenamefont {Zheng},\ and\
  \citenamefont {Ishida}}]{ASAYAMA1996221}%
  \BibitemOpen
  \bibfield  {author} {\bibinfo {author} {\bibfnamefont {K.}~\bibnamefont
  {Asayama}}, \bibinfo {author} {\bibfnamefont {Y.}~\bibnamefont {Kitaoka}},
  \bibinfo {author} {\bibfnamefont {G.-q.}\ \bibnamefont {Zheng}},\ and\
  \bibinfo {author} {\bibfnamefont {K.}~\bibnamefont {Ishida}},\ }\bibfield
  {title} {\bibinfo {title} {{NMR studies of high Tc superconductors}},\ }\href
  {https://doi.org/https://doi.org/10.1016/0079-6565(95)01025-4} {\bibfield
  {journal} {\bibinfo  {journal} {Progress in Nuclear Magnetic Resonance
  Spectroscopy}\ }\textbf {\bibinfo {volume} {28}},\ \bibinfo {pages} {221}
  (\bibinfo {year} {1996})}\BibitemShut {NoStop}%
\bibitem [{\citenamefont {Imai}\ \emph {et~al.}(1995)\citenamefont {Imai},
  \citenamefont {Slichter}, \citenamefont {Cobb},\ and\ \citenamefont
  {Markert}}]{Imai1995}%
  \BibitemOpen
  \bibfield  {author} {\bibinfo {author} {\bibfnamefont {T.}~\bibnamefont
  {Imai}}, \bibinfo {author} {\bibfnamefont {C.~P.}\ \bibnamefont {Slichter}},
  \bibinfo {author} {\bibfnamefont {J.~L.}\ \bibnamefont {Cobb}},\ and\
  \bibinfo {author} {\bibfnamefont {J.~T.}\ \bibnamefont {Markert}},\
  }\bibfield  {title} {\bibinfo {title} {{Superconductivity and spin
  fluctuations in the electron-doped infinitely-layered high Tc superconductor
  Sr$_{0.9}$La$_{0.1}$CuO$_{2}$ ($T_{\rm c} = 42$ K)}},\ }\href
  {https://doi.org/https://doi.org/10.1016/0022-3697(95)00226-X} {\bibfield
  {journal} {\bibinfo  {journal} {Journal of Physics and Chemistry of Solids}\
  }\textbf {\bibinfo {volume} {56}},\ \bibinfo {pages} {1921} (\bibinfo {year}
  {1995})}\BibitemShut {NoStop}%
\bibitem [{\citenamefont {Rigamonti}\ \emph {et~al.}(1998)\citenamefont
  {Rigamonti}, \citenamefont {Borsa},\ and\ \citenamefont
  {Carretta}}]{rigamonti_1998}%
  \BibitemOpen
  \bibfield  {author} {\bibinfo {author} {\bibfnamefont {A.}~\bibnamefont
  {Rigamonti}}, \bibinfo {author} {\bibfnamefont {F.}~\bibnamefont {Borsa}},\
  and\ \bibinfo {author} {\bibfnamefont {P.}~\bibnamefont {Carretta}},\
  }\bibfield  {title} {\bibinfo {title} {Basic aspects and main results of
  nmr-nqr spectroscopies in high-temperature superconductors},\ }\href
  {https://doi.org/10.1088/0034-4885/61/10/002} {\bibfield  {journal} {\bibinfo
   {journal} {Reports on Progress in Physics}\ }\textbf {\bibinfo {volume}
  {61}},\ \bibinfo {pages} {1367} (\bibinfo {year} {1998})}\BibitemShut
  {NoStop}%
\bibitem [{\citenamefont {Jurkutat}\ \emph {et~al.}(2019)\citenamefont
  {Jurkutat}, \citenamefont {Avramovska}, \citenamefont {Williams},
  \citenamefont {Dernbach}, \citenamefont {Pavicevic},\ and\ \citenamefont
  {Haase}}]{Jurkutat2019}%
  \BibitemOpen
  \bibfield  {author} {\bibinfo {author} {\bibfnamefont {M.}~\bibnamefont
  {Jurkutat}}, \bibinfo {author} {\bibfnamefont {M.}~\bibnamefont
  {Avramovska}}, \bibinfo {author} {\bibfnamefont {G.~V.~M.}\ \bibnamefont
  {Williams}}, \bibinfo {author} {\bibfnamefont {D.}~\bibnamefont {Dernbach}},
  \bibinfo {author} {\bibfnamefont {D.}~\bibnamefont {Pavicevic}},\ and\
  \bibinfo {author} {\bibfnamefont {J.}~\bibnamefont {Haase}},\ }\bibfield
  {title} {\bibinfo {title} {{Phenomenology of $^{63}$Cu Nuclear Relaxation in
  Cuprate Superconductors}},\ }\href
  {https://doi.org/10.1007/s10948-019-05275-6} {\bibfield  {journal} {\bibinfo
  {journal} {Journal of Superconductivity and Novel Magnetism}\ }\textbf
  {\bibinfo {volume} {32}},\ \bibinfo {pages} {3369} (\bibinfo {year}
  {2019})}\BibitemShut {NoStop}%
\bibitem [{\citenamefont {Hebel}(1959)}]{Hebel1959}%
  \BibitemOpen
  \bibfield  {author} {\bibinfo {author} {\bibfnamefont {L.~C.}\ \bibnamefont
  {Hebel}},\ }\bibfield  {title} {\bibinfo {title} {{Theory of Nuclear Spin
  Relaxation in Superconductors}},\ }\href
  {https://doi.org/10.1103/PhysRev.116.79} {\bibfield  {journal} {\bibinfo
  {journal} {Physical Review}\ }\textbf {\bibinfo {volume} {116}},\ \bibinfo
  {pages} {79} (\bibinfo {year} {1959})}\BibitemShut {NoStop}%
\bibitem [{\citenamefont {Slichter}(1990)}]{Slichter1990}%
  \BibitemOpen
  \bibfield  {author} {\bibinfo {author} {\bibfnamefont {C.~P.}\ \bibnamefont
  {Slichter}},\ }\href {https://doi.org/10.1007/978-3-662-09441-9} {\emph
  {\bibinfo {title} {Principles of Magnetic Resonance}}},\ Springer Series in
  Solid-State Sciences\ (\bibinfo  {publisher} {Springer Berlin Heidelberg},\
  \bibinfo {year} {1990})\BibitemShut {NoStop}%
\bibitem [{\citenamefont {Slichter}(2007)}]{Slichter2017}%
  \BibitemOpen
  \bibfield  {author} {\bibinfo {author} {\bibfnamefont {C.~P.}\ \bibnamefont
  {Slichter}},\ }\bibinfo {title} {Magnetic resonance studies of high
  temperature superconductors},\ in\ \href
  {https://doi.org/10.1007/978-0-387-68734-6_5} {\emph {\bibinfo {booktitle}
  {Handbook of High-Temperature Superconductivity: Theory and Experiment}}},\
  \bibinfo {editor} {edited by\ \bibinfo {editor} {\bibfnamefont {J.~R.}\
  \bibnamefont {Schrieffer}}\ and\ \bibinfo {editor} {\bibfnamefont {J.~S.}\
  \bibnamefont {Brooks}}}\ (\bibinfo  {publisher} {Springer New York},\
  \bibinfo {address} {New York, NY},\ \bibinfo {year} {2007})\ pp.\ \bibinfo
  {pages} {215--256}\BibitemShut {NoStop}%
\bibitem [{\citenamefont {MacLaughlin}(1976)}]{MacLaughlinDE}%
  \BibitemOpen
  \bibfield  {author} {\bibinfo {author} {\bibfnamefont {D.~E.}\ \bibnamefont
  {MacLaughlin}},\ }\bibinfo {title} {Magnetic resonance in the superconducting
  state},\ in\ \href {https://doi.org/10.1016/s0081-1947(08)60541-x} {\emph
  {\bibinfo {booktitle} {Solid State Physics}}},\ Vol.~\bibinfo {volume} {31},\
  \bibinfo {editor} {edited by\ \bibinfo {editor} {\bibfnamefont {F.~S.}\
  \bibnamefont {Henry~Ehrenreich}}\ and\ \bibinfo {editor} {\bibfnamefont
  {T.}~\bibnamefont {David}}}\ (\bibinfo  {publisher} {Academic Press},\
  \bibinfo {year} {1976})\ pp.\ \bibinfo {pages} {1--69}\BibitemShut {NoStop}%
\bibitem [{\citenamefont {Wada}\ and\ \citenamefont
  {Asayama}(1973)}]{Wada1973}%
  \BibitemOpen
  \bibfield  {author} {\bibinfo {author} {\bibfnamefont {S.}~\bibnamefont
  {Wada}}\ and\ \bibinfo {author} {\bibfnamefont {K.}~\bibnamefont {Asayama}},\
  }\bibfield  {title} {\bibinfo {title} {{Nuclear Quadrupole Spin-Lattice
  Relaxation of Ta$^{181}$ in Type II Superconducting Ta$_3$Sn}},\ }\href
  {https://doi.org/10.1143/jpsj.34.1168} {\bibfield  {journal} {\bibinfo
  {journal} {Journal of the Physical Society of Japan}\ }\textbf {\bibinfo
  {volume} {34}},\ \bibinfo {pages} {1168} (\bibinfo {year}
  {1973})}\BibitemShut {NoStop}%
\bibitem [{\citenamefont {Li}\ \emph {et~al.}(2016)\citenamefont {Li},
  \citenamefont {Jiao}, \citenamefont {Cao},\ and\ \citenamefont
  {Zheng}}]{Li2016TaPdTe}%
  \BibitemOpen
  \bibfield  {author} {\bibinfo {author} {\bibfnamefont {Z.}~\bibnamefont
  {Li}}, \bibinfo {author} {\bibfnamefont {W.~H.}\ \bibnamefont {Jiao}},
  \bibinfo {author} {\bibfnamefont {G.~H.}\ \bibnamefont {Cao}},\ and\ \bibinfo
  {author} {\bibfnamefont {G.-q.}\ \bibnamefont {Zheng}},\ }\bibfield  {title}
  {\bibinfo {title} {{Charge fluctuations and nodeless superconductivity in
  quasi-one-dimensional Ta$_{4}$Pd$_{3}$Te$_{16}$ revealed by $^{125}$Te-NMR
  and $^{181}$Ta-NQR}},\ }\href {https://doi.org/10.1103/PhysRevB.94.174511}
  {\bibfield  {journal} {\bibinfo  {journal} {Phys. Rev. B}\ }\textbf {\bibinfo
  {volume} {94}},\ \bibinfo {pages} {174511} (\bibinfo {year}
  {2016})}\BibitemShut {NoStop}%
\bibitem [{\citenamefont {Xiang}\ and\ \citenamefont {Wu}(2022)}]{Xiang2022}%
  \BibitemOpen
  \bibfield  {author} {\bibinfo {author} {\bibfnamefont {T.}~\bibnamefont
  {Xiang}}\ and\ \bibinfo {author} {\bibfnamefont {C.}~\bibnamefont {Wu}},\
  }\href
  {https://www.cambridge.org/core/product/196ECDAD19E69C11438E81631C12198C}
  {\emph {\bibinfo {title} {D-wave Superconductivity}}}\ (\bibinfo  {publisher}
  {Cambridge University Press},\ \bibinfo {address} {Cambridge},\ \bibinfo
  {year} {2022})\BibitemShut {NoStop}%
\bibitem [{\citenamefont {Karpinski}\ \emph {et~al.}(1988)\citenamefont
  {Karpinski}, \citenamefont {Kaldis}, \citenamefont {Jilek}, \citenamefont
  {Rusiecki},\ and\ \citenamefont {Bucher}}]{Karpinski1988}%
  \BibitemOpen
  \bibfield  {author} {\bibinfo {author} {\bibfnamefont {J.}~\bibnamefont
  {Karpinski}}, \bibinfo {author} {\bibfnamefont {E.}~\bibnamefont {Kaldis}},
  \bibinfo {author} {\bibfnamefont {E.}~\bibnamefont {Jilek}}, \bibinfo
  {author} {\bibfnamefont {S.}~\bibnamefont {Rusiecki}},\ and\ \bibinfo
  {author} {\bibfnamefont {B.}~\bibnamefont {Bucher}},\ }\bibfield  {title}
  {\bibinfo {title} {{Bulk synthesis of the 81-K superconductor
  YBa$_2$Cu$_4$O$_8$ at high oxygen pressure}},\ }\href
  {https://doi.org/10.1038/336660a0} {\bibfield  {journal} {\bibinfo  {journal}
  {Nature}\ }\textbf {\bibinfo {volume} {336}},\ \bibinfo {pages} {660}
  (\bibinfo {year} {1988})}\BibitemShut {NoStop}%
\bibitem [{\citenamefont {Raffa}\ \emph {et~al.}(1998)\citenamefont {Raffa},
  \citenamefont {Ohno}, \citenamefont {Mali}, \citenamefont {Roos},
  \citenamefont {Brinkmann}, \citenamefont {Conder},\ and\ \citenamefont
  {Eremin}}]{Raffa1998}%
  \BibitemOpen
  \bibfield  {author} {\bibinfo {author} {\bibfnamefont {F.}~\bibnamefont
  {Raffa}}, \bibinfo {author} {\bibfnamefont {T.}~\bibnamefont {Ohno}},
  \bibinfo {author} {\bibfnamefont {M.}~\bibnamefont {Mali}}, \bibinfo {author}
  {\bibfnamefont {J.}~\bibnamefont {Roos}}, \bibinfo {author} {\bibfnamefont
  {D.}~\bibnamefont {Brinkmann}}, \bibinfo {author} {\bibfnamefont
  {K.}~\bibnamefont {Conder}},\ and\ \bibinfo {author} {\bibfnamefont
  {M.}~\bibnamefont {Eremin}},\ }\bibfield  {title} {\bibinfo {title} {{Isotope
  Dependence of the Spin Gap in YBa$_2$Cu$_4$O$_8$ as Determined by Cu NQR
  Relaxation}},\ }\href {https://doi.org/10.1103/PhysRevLett.81.5912}
  {\bibfield  {journal} {\bibinfo  {journal} {Physical Review Letters}\
  }\textbf {\bibinfo {volume} {81}},\ \bibinfo {pages} {5912} (\bibinfo {year}
  {1998})}\BibitemShut {NoStop}%
\bibitem [{\citenamefont {Suter}\ \emph {et~al.}(1998)\citenamefont {Suter},
  \citenamefont {Mali}, \citenamefont {Roos},\ and\ \citenamefont
  {Brinkmann}}]{Suter1998}%
  \BibitemOpen
  \bibfield  {author} {\bibinfo {author} {\bibfnamefont {A.}~\bibnamefont
  {Suter}}, \bibinfo {author} {\bibfnamefont {M.}~\bibnamefont {Mali}},
  \bibinfo {author} {\bibfnamefont {J.}~\bibnamefont {Roos}},\ and\ \bibinfo
  {author} {\bibfnamefont {D.}~\bibnamefont {Brinkmann}},\ }\bibfield  {title}
  {\bibinfo {title} {{Mixed magnetic and quadrupolar relaxation in the presence
  of a dominant static Zeeman Hamiltonian}},\ }\href
  {https://doi.org/10.1088/0953-8984/10/26/022} {\bibfield  {journal} {\bibinfo
   {journal} {Journal of Physics-Condensed Matter}\ }\textbf {\bibinfo {volume}
  {10}},\ \bibinfo {pages} {5977} (\bibinfo {year} {1998})}\BibitemShut
  {NoStop}%
\bibitem [{\citenamefont {Mangelschots}\ \emph {et~al.}(1992)\citenamefont
  {Mangelschots}, \citenamefont {Mali}, \citenamefont {Roos}, \citenamefont
  {Brinkmann}, \citenamefont {Rusiecki}, \citenamefont {Karpinski},\ and\
  \citenamefont {Kaldis}}]{Mangelschots1992}%
  \BibitemOpen
  \bibfield  {author} {\bibinfo {author} {\bibfnamefont {I.}~\bibnamefont
  {Mangelschots}}, \bibinfo {author} {\bibfnamefont {M.}~\bibnamefont {Mali}},
  \bibinfo {author} {\bibfnamefont {J.}~\bibnamefont {Roos}}, \bibinfo {author}
  {\bibfnamefont {D.}~\bibnamefont {Brinkmann}}, \bibinfo {author}
  {\bibfnamefont {S.}~\bibnamefont {Rusiecki}}, \bibinfo {author}
  {\bibfnamefont {J.}~\bibnamefont {Karpinski}},\ and\ \bibinfo {author}
  {\bibfnamefont {E.}~\bibnamefont {Kaldis}},\ }\bibfield  {title} {\bibinfo
  {title} {{$^{17}$O NMR study in aligned YBa$_2$Cu$_4$O$_8$ powder}},\ }\href
  {https://doi.org/https://doi.org/10.1016/S0921-4534(05)80005-X} {\bibfield
  {journal} {\bibinfo  {journal} {Physica C: Superconductivity}\ }\textbf
  {\bibinfo {volume} {194}},\ \bibinfo {pages} {277} (\bibinfo {year}
  {1992})}\BibitemShut {NoStop}%
\bibitem [{\citenamefont {Suter}\ \emph {et~al.}(2000)\citenamefont {Suter},
  \citenamefont {Mali}, \citenamefont {Roos},\ and\ \citenamefont
  {Brinkmann}}]{Suter2000}%
  \BibitemOpen
  \bibfield  {author} {\bibinfo {author} {\bibfnamefont {A.}~\bibnamefont
  {Suter}}, \bibinfo {author} {\bibfnamefont {M.}~\bibnamefont {Mali}},
  \bibinfo {author} {\bibfnamefont {J.}~\bibnamefont {Roos}},\ and\ \bibinfo
  {author} {\bibfnamefont {D.}~\bibnamefont {Brinkmann}},\ }\bibfield  {title}
  {\bibinfo {title} {{Charge degree of freedom and the single-spin fluid model
  in YBa$_2$Cu$_4$O$_8$}},\ }\href
  {https://doi.org/10.1103/PhysRevLett.84.4938} {\bibfield  {journal} {\bibinfo
   {journal} {Physical Review Letters}\ }\textbf {\bibinfo {volume} {84}},\
  \bibinfo {pages} {4938} (\bibinfo {year} {2000})}\BibitemShut {NoStop}%
\bibitem [{\citenamefont {Goto}\ \emph {et~al.}(1998)\citenamefont {Goto},
  \citenamefont {Shimizu}, \citenamefont {Aoki}, \citenamefont {Kato},
  \citenamefont {Yoshimura}, \citenamefont {Kosuge}, \citenamefont
  {Matsumoto},\ and\ \citenamefont {Yamada}}]{Goto1998}%
  \BibitemOpen
  \bibfield  {author} {\bibinfo {author} {\bibfnamefont {A.}~\bibnamefont
  {Goto}}, \bibinfo {author} {\bibfnamefont {T.}~\bibnamefont {Shimizu}},
  \bibinfo {author} {\bibfnamefont {H.}~\bibnamefont {Aoki}}, \bibinfo {author}
  {\bibfnamefont {M.}~\bibnamefont {Kato}}, \bibinfo {author} {\bibfnamefont
  {K.}~\bibnamefont {Yoshimura}}, \bibinfo {author} {\bibfnamefont
  {K.}~\bibnamefont {Kosuge}}, \bibinfo {author} {\bibfnamefont
  {T.}~\bibnamefont {Matsumoto}},\ and\ \bibinfo {author} {\bibfnamefont
  {Y.}~\bibnamefont {Yamada}},\ }\bibfield  {title} {\bibinfo {title}
  {{Anisotropy Study of the Spin-Lattice Relaxation Rates at the Cu(1) Chain
  Sites of YBa$_2$Cu$_3$O$_7$ and YBa$_2$Cu$_4$O$_8$}},\ }\href
  {https://doi.org/10.1143/JPSJ.67.759} {\bibfield  {journal} {\bibinfo
  {journal} {Journal of the Physical Society of Japan}\ }\textbf {\bibinfo
  {volume} {67}},\ \bibinfo {pages} {759} (\bibinfo {year} {1998})}\BibitemShut
  {NoStop}%
\bibitem [{\citenamefont {Raffa}\ \emph {et~al.}(1999)\citenamefont {Raffa},
  \citenamefont {Mali}, \citenamefont {Suter}, \citenamefont {Zavidonov},
  \citenamefont {Roos}, \citenamefont {Brinkmann},\ and\ \citenamefont
  {Conder}}]{Raffa1999}%
  \BibitemOpen
  \bibfield  {author} {\bibinfo {author} {\bibfnamefont {F.}~\bibnamefont
  {Raffa}}, \bibinfo {author} {\bibfnamefont {M.}~\bibnamefont {Mali}},
  \bibinfo {author} {\bibfnamefont {A.}~\bibnamefont {Suter}}, \bibinfo
  {author} {\bibfnamefont {A.~Y.}\ \bibnamefont {Zavidonov}}, \bibinfo {author}
  {\bibfnamefont {J.}~\bibnamefont {Roos}}, \bibinfo {author} {\bibfnamefont
  {D.}~\bibnamefont {Brinkmann}},\ and\ \bibinfo {author} {\bibfnamefont
  {K.}~\bibnamefont {Conder}},\ }\bibfield  {title} {\bibinfo {title} {{Spin
  and charge dynamics in the Cu-O chains of
  ${\mathrm{YBa}}_{2}{\mathrm{Cu}}_{4}{\mathrm{O}}_{8}$}},\ }\href
  {https://doi.org/10.1103/PhysRevB.60.3636} {\bibfield  {journal} {\bibinfo
  {journal} {Physical Review B}\ }\textbf {\bibinfo {volume} {60}},\ \bibinfo
  {pages} {3636} (\bibinfo {year} {1999})}\BibitemShut {NoStop}%
\bibitem [{\citenamefont {Xiang}\ and\ \citenamefont
  {Wheatley}(1996)}]{Xiang1996}%
  \BibitemOpen
  \bibfield  {author} {\bibinfo {author} {\bibfnamefont {T.}~\bibnamefont
  {Xiang}}\ and\ \bibinfo {author} {\bibfnamefont {J.~M.}\ \bibnamefont
  {Wheatley}},\ }\bibfield  {title} {\bibinfo {title} {{Superfluid Anisotropy
  in YBCO: Evidence for Pair Tunneling Superconductivity}},\ }\href
  {https://doi.org/10.1103/PhysRevLett.76.134} {\bibfield  {journal} {\bibinfo
  {journal} {Physical Review Letters}\ }\textbf {\bibinfo {volume} {76}},\
  \bibinfo {pages} {134} (\bibinfo {year} {1996})}\BibitemShut {NoStop}%
\bibitem [{\citenamefont {Serafin}\ \emph {et~al.}(2010)\citenamefont
  {Serafin}, \citenamefont {Fletcher}, \citenamefont {Adachi}, \citenamefont
  {Hussey},\ and\ \citenamefont {Carrington}}]{Serafin2010}%
  \BibitemOpen
  \bibfield  {author} {\bibinfo {author} {\bibfnamefont {A.}~\bibnamefont
  {Serafin}}, \bibinfo {author} {\bibfnamefont {J.~D.}\ \bibnamefont
  {Fletcher}}, \bibinfo {author} {\bibfnamefont {S.}~\bibnamefont {Adachi}},
  \bibinfo {author} {\bibfnamefont {N.~E.}\ \bibnamefont {Hussey}},\ and\
  \bibinfo {author} {\bibfnamefont {A.}~\bibnamefont {Carrington}},\ }\bibfield
   {title} {\bibinfo {title} {{Destruction of chain superconductivity in
  ${\text{YBa}}_{2}{\text{Cu}}_{4}{\text{O}}_{8}$ in a weak magnetic field}},\
  }\href {https://doi.org/10.1103/PhysRevB.82.140506} {\bibfield  {journal}
  {\bibinfo  {journal} {Physical Review B}\ }\textbf {\bibinfo {volume} {82}},\
  \bibinfo {pages} {140506} (\bibinfo {year} {2010})}\BibitemShut {NoStop}%
\bibitem [{\citenamefont {Atkinson}\ and\ \citenamefont
  {Carbotte}(1995)}]{Atkinson1995}%
  \BibitemOpen
  \bibfield  {author} {\bibinfo {author} {\bibfnamefont {W.~A.}\ \bibnamefont
  {Atkinson}}\ and\ \bibinfo {author} {\bibfnamefont {J.~P.}\ \bibnamefont
  {Carbotte}},\ }\bibfield  {title} {\bibinfo {title} {{Effect of proximity
  coupling of chains and planes on the penetration-depth anisotropy in
  ${\mathrm{YBa}}_{2}$${\mathrm{Cu}}_{3}$${\mathrm{O}}_{7}$}},\ }\href
  {https://doi.org/10.1103/PhysRevB.52.10601} {\bibfield  {journal} {\bibinfo
  {journal} {Physical Review B}\ }\textbf {\bibinfo {volume} {52}},\ \bibinfo
  {pages} {10601} (\bibinfo {year} {1995})}\BibitemShut {NoStop}%
\bibitem [{\citenamefont {Gagnon}\ \emph {et~al.}(1997)\citenamefont {Gagnon},
  \citenamefont {Pu}, \citenamefont {Ellman},\ and\ \citenamefont
  {Taillefer}}]{Gagnon1997}%
  \BibitemOpen
  \bibfield  {author} {\bibinfo {author} {\bibfnamefont {R.}~\bibnamefont
  {Gagnon}}, \bibinfo {author} {\bibfnamefont {S.}~\bibnamefont {Pu}}, \bibinfo
  {author} {\bibfnamefont {B.}~\bibnamefont {Ellman}},\ and\ \bibinfo {author}
  {\bibfnamefont {L.}~\bibnamefont {Taillefer}},\ }\bibfield  {title} {\bibinfo
  {title} {{Anisotropy of Heat Conduction in
  ${\mathrm{YBa}}_{2}{\mathrm{Cu}}_{3}{\mathrm{O}}_{6.9}$: A Probe of Chain
  Superconductivity}},\ }\href {https://doi.org/10.1103/PhysRevLett.78.1976}
  {\bibfield  {journal} {\bibinfo  {journal} {Physical Review Letters}\
  }\textbf {\bibinfo {volume} {78}},\ \bibinfo {pages} {1976} (\bibinfo {year}
  {1997})}\BibitemShut {NoStop}%
\bibitem [{\citenamefont {Fong}\ \emph {et~al.}(1995)\citenamefont {Fong},
  \citenamefont {Keimer}, \citenamefont {Anderson}, \citenamefont {Reznik},
  \citenamefont {Doğan},\ and\ \citenamefont {Aksay}}]{Fong1995}%
  \BibitemOpen
  \bibfield  {author} {\bibinfo {author} {\bibfnamefont {H.~F.}\ \bibnamefont
  {Fong}}, \bibinfo {author} {\bibfnamefont {B.}~\bibnamefont {Keimer}},
  \bibinfo {author} {\bibfnamefont {P.~W.}\ \bibnamefont {Anderson}}, \bibinfo
  {author} {\bibfnamefont {D.}~\bibnamefont {Reznik}}, \bibinfo {author}
  {\bibfnamefont {F.}~\bibnamefont {Doğan}},\ and\ \bibinfo {author}
  {\bibfnamefont {I.~A.}\ \bibnamefont {Aksay}},\ }\bibfield  {title} {\bibinfo
  {title} {{Phonon and Magnetic Neutron Scattering at 41 meV in
  YBa$_2$Cu$_3$O$_7$}},\ }\href {https://doi.org/10.1103/PhysRevLett.75.316}
  {\bibfield  {journal} {\bibinfo  {journal} {Physical Review Letters}\
  }\textbf {\bibinfo {volume} {75}},\ \bibinfo {pages} {316} (\bibinfo {year}
  {1995})}\BibitemShut {NoStop}%
\bibitem [{\citenamefont {Koyama}\ and\ \citenamefont
  {Tachiki}(1989)}]{koyama_PhysRevB.39.2279}%
  \BibitemOpen
  \bibfield  {author} {\bibinfo {author} {\bibfnamefont {T.}~\bibnamefont
  {Koyama}}\ and\ \bibinfo {author} {\bibfnamefont {M.}~\bibnamefont
  {Tachiki}},\ }\bibfield  {title} {\bibinfo {title} {Theory of nuclear
  relaxation in superconducting high-${T}_{c}$ oxides},\ }\href
  {https://doi.org/10.1103/PhysRevB.39.2279} {\bibfield  {journal} {\bibinfo
  {journal} {Phys. Rev. B}\ }\textbf {\bibinfo {volume} {39}},\ \bibinfo
  {pages} {2279} (\bibinfo {year} {1989})}\BibitemShut {NoStop}%
\bibitem [{\citenamefont {Monien}\ and\ \citenamefont
  {Pines}(1990)}]{pines1990}%
  \BibitemOpen
  \bibfield  {author} {\bibinfo {author} {\bibfnamefont {H.}~\bibnamefont
  {Monien}}\ and\ \bibinfo {author} {\bibfnamefont {D.}~\bibnamefont {Pines}},\
  }\bibfield  {title} {\bibinfo {title} {{Spin excitations and pairing gaps in
  the superconducting state of YBa$_{2}$Cu$_{3}$O$_{7-\delta}$}},\ }\href
  {https://doi.org/10.1103/PhysRevB.41.6297} {\bibfield  {journal} {\bibinfo
  {journal} {Phys. Rev. B}\ }\textbf {\bibinfo {volume} {41}},\ \bibinfo
  {pages} {6297} (\bibinfo {year} {1990})}\BibitemShut {NoStop}%
\bibitem [{\citenamefont {Thelen}\ \emph {et~al.}(1993)\citenamefont {Thelen},
  \citenamefont {Pines},\ and\ \citenamefont {Lu}}]{pines1993}%
  \BibitemOpen
  \bibfield  {author} {\bibinfo {author} {\bibfnamefont {D.}~\bibnamefont
  {Thelen}}, \bibinfo {author} {\bibfnamefont {D.}~\bibnamefont {Pines}},\ and\
  \bibinfo {author} {\bibfnamefont {J.~P.}\ \bibnamefont {Lu}},\ }\bibfield
  {title} {\bibinfo {title} {{Evidence for
  ${\mathit{d}}_{{\mathit{x}}^{2}-{\mathit{y}}^{2}}$ pairing from
  nuclear-magnetic-resonance experiments in the superconducting state of
  ${\mathrm{YBa}}_{2}$${\mathrm{Cu}}_{3}$${\mathrm{O}}_{7}$}},\ }\href
  {https://doi.org/10.1103/PhysRevB.47.9151} {\bibfield  {journal} {\bibinfo
  {journal} {Phys. Rev. B}\ }\textbf {\bibinfo {volume} {47}},\ \bibinfo
  {pages} {9151} (\bibinfo {year} {1993})}\BibitemShut {NoStop}%
\bibitem [{\citenamefont {Zoli}(1991)}]{zoli1991}%
  \BibitemOpen
  \bibfield  {author} {\bibinfo {author} {\bibfnamefont {M.}~\bibnamefont
  {Zoli}},\ }\bibfield  {title} {\bibinfo {title} {{Smearing of the
  Hebel-Slichter Peak by 2D Fluctuationsin Layered Cuprate Superconductors}},\
  }\href {https://doi.org/10.1143/JPSJ.60.3837} {\bibfield  {journal} {\bibinfo
   {journal} {Journal of the Physical Society of Japan}\ }\textbf {\bibinfo
  {volume} {60}},\ \bibinfo {pages} {3837} (\bibinfo {year}
  {1991})}\BibitemShut {NoStop}%
\bibitem [{\citenamefont {Statt}(1990)}]{statt1990}%
  \BibitemOpen
  \bibfield  {author} {\bibinfo {author} {\bibfnamefont {B.~W.}\ \bibnamefont
  {Statt}},\ }\bibfield  {title} {\bibinfo {title} {{Anisotropic gap and
  quasiparticle-damping effects on NMR measurements of high-temperature
  superconductors}},\ }\href {https://doi.org/10.1103/PhysRevB.42.6805}
  {\bibfield  {journal} {\bibinfo  {journal} {Physical Review B}\ }\textbf
  {\bibinfo {volume} {42}},\ \bibinfo {pages} {6805} (\bibinfo {year}
  {1990})}\BibitemShut {NoStop}%
\bibitem [{\citenamefont {Emery}\ and\ \citenamefont
  {Kivelson}(1995)}]{Emery1995}%
  \BibitemOpen
  \bibfield  {author} {\bibinfo {author} {\bibfnamefont {V.~J.}\ \bibnamefont
  {Emery}}\ and\ \bibinfo {author} {\bibfnamefont {S.~A.}\ \bibnamefont
  {Kivelson}},\ }\bibfield  {title} {\bibinfo {title} {Importance of phase
  fluctuations in superconductors with small superfluid density},\ }\href
  {https://doi.org/10.1038/374434a0} {\bibfield  {journal} {\bibinfo  {journal}
  {Nature}\ }\textbf {\bibinfo {volume} {374}},\ \bibinfo {pages} {434}
  (\bibinfo {year} {1995})}\BibitemShut {NoStop}%
\bibitem [{\citenamefont {Keimer}\ \emph {et~al.}(2015)\citenamefont {Keimer},
  \citenamefont {Kivelson}, \citenamefont {Norman}, \citenamefont {Uchida},\
  and\ \citenamefont {Zaanen}}]{Keimer2015}%
  \BibitemOpen
  \bibfield  {author} {\bibinfo {author} {\bibfnamefont {B.}~\bibnamefont
  {Keimer}}, \bibinfo {author} {\bibfnamefont {S.~A.}\ \bibnamefont
  {Kivelson}}, \bibinfo {author} {\bibfnamefont {M.~R.}\ \bibnamefont
  {Norman}}, \bibinfo {author} {\bibfnamefont {S.}~\bibnamefont {Uchida}},\
  and\ \bibinfo {author} {\bibfnamefont {J.}~\bibnamefont {Zaanen}},\
  }\bibfield  {title} {\bibinfo {title} {{From quantum matter to
  high-temperature superconductivity in copper oxides}},\ }\href
  {https://doi.org/10.1038/nature14165} {\bibfield  {journal} {\bibinfo
  {journal} {Nature}\ }\textbf {\bibinfo {volume} {518}},\ \bibinfo {pages}
  {179} (\bibinfo {year} {2015})}\BibitemShut {NoStop}%
\bibitem [{\citenamefont {Wu}\ \emph {et~al.}(2011)\citenamefont {Wu},
  \citenamefont {Mayaffre}, \citenamefont {Kramer}, \citenamefont {Horvatic},
  \citenamefont {Berthier}, \citenamefont {Hardy}, \citenamefont {Liang},
  \citenamefont {Bonn},\ and\ \citenamefont {Julien}}]{Wu2011}%
  \BibitemOpen
  \bibfield  {author} {\bibinfo {author} {\bibfnamefont {T.}~\bibnamefont
  {Wu}}, \bibinfo {author} {\bibfnamefont {H.}~\bibnamefont {Mayaffre}},
  \bibinfo {author} {\bibfnamefont {S.}~\bibnamefont {Kramer}}, \bibinfo
  {author} {\bibfnamefont {M.}~\bibnamefont {Horvatic}}, \bibinfo {author}
  {\bibfnamefont {C.}~\bibnamefont {Berthier}}, \bibinfo {author}
  {\bibfnamefont {W.~N.}\ \bibnamefont {Hardy}}, \bibinfo {author}
  {\bibfnamefont {R.~X.}\ \bibnamefont {Liang}}, \bibinfo {author}
  {\bibfnamefont {D.~A.}\ \bibnamefont {Bonn}},\ and\ \bibinfo {author}
  {\bibfnamefont {M.~H.}\ \bibnamefont {Julien}},\ }\bibfield  {title}
  {\bibinfo {title} {{Magnetic-field-induced charge-stripe order in the
  high-temperature superconductor YBa$_2$Cu$_3$O$_y$}},\ }\href
  {https://doi.org/10.1038/nature10345} {\bibfield  {journal} {\bibinfo
  {journal} {Nature}\ }\textbf {\bibinfo {volume} {477}},\ \bibinfo {pages}
  {191} (\bibinfo {year} {2011})}\BibitemShut {NoStop}%
\bibitem [{\citenamefont {Oguchi}\ \emph {et~al.}(1990)\citenamefont {Oguchi},
  \citenamefont {Sasaki},\ and\ \citenamefont {Terakura}}]{Oguchi1990}%
  \BibitemOpen
  \bibfield  {author} {\bibinfo {author} {\bibfnamefont {T.}~\bibnamefont
  {Oguchi}}, \bibinfo {author} {\bibfnamefont {T.}~\bibnamefont {Sasaki}},\
  and\ \bibinfo {author} {\bibfnamefont {K.}~\bibnamefont {Terakura}},\
  }\bibfield  {title} {\bibinfo {title} {{Electronic band structure of
  YBa$_2$Cu$_4$O$_8$}},\ }\href
  {https://doi.org/https://doi.org/10.1016/0921-4534(90)90617-N} {\bibfield
  {journal} {\bibinfo  {journal} {Physica C: Superconductivity}\ }\textbf
  {\bibinfo {volume} {172}},\ \bibinfo {pages} {277} (\bibinfo {year}
  {1990})}\BibitemShut {NoStop}%
\bibitem [{\citenamefont {Li}\ \emph {et~al.}(2011)\citenamefont {Li},
  \citenamefont {Sun}, \citenamefont {Lin}, \citenamefont {Su}, \citenamefont
  {Hu},\ and\ \citenamefont {Zheng}}]{LZ2011}%
  \BibitemOpen
  \bibfield  {author} {\bibinfo {author} {\bibfnamefont {Z.}~\bibnamefont
  {Li}}, \bibinfo {author} {\bibfnamefont {D.~L.}\ \bibnamefont {Sun}},
  \bibinfo {author} {\bibfnamefont {C.~T.}\ \bibnamefont {Lin}}, \bibinfo
  {author} {\bibfnamefont {Y.~H.}\ \bibnamefont {Su}}, \bibinfo {author}
  {\bibfnamefont {J.~P.}\ \bibnamefont {Hu}},\ and\ \bibinfo {author}
  {\bibfnamefont {G.~Q.}\ \bibnamefont {Zheng}},\ }\bibfield  {title} {\bibinfo
  {title} {{Nodeless energy gaps of single-crystalline
  Ba$_{0.68}$K$_{0.32}$Fe$_2$As$_2$ as seen via $^{75}$As NMR}},\ }\href
  {https://doi.org/10.1103/PhysRevB.83.140506} {\bibfield  {journal} {\bibinfo
  {journal} {Physical Review B}\ }\textbf {\bibinfo {volume} {83}},\ \bibinfo
  {pages} {140506} (\bibinfo {year} {2011})}\BibitemShut {NoStop}%
\end{thebibliography}
\end{document}